\documentclass[12pt]{iopart}
\usepackage{iopams}
\usepackage{graphicx}

\newcommand{\R}{\mathbb{R}}
\newcommand{\C}{\mathbb{C}}
\newcommand{\Z}{\mathbb{Z}}
\newcommand{\Ad}{\mathrm{Ad}}
\newcommand{\ad}{\mathrm{ad}}

\begin{document}

{\normalsize \hfill ITP-UU-09/24}\\
\vspace{-1.5cm}
{\normalsize \hfill SPIN-09/22}\\
${}$\\

\title{In Search of Fundamental Discreteness in  2+1 Dimensional Quantum Gravity}

\author{T.G. Budd and R. Loll}

\address{Institute for Theoretical Physics, Utrecht University, Postbus 80195, 3508 TD Utrecht, The Netherlands}
\ead{t.g.budd@uu.nl, r.loll@uu.nl}
\begin{abstract}
Inspired by previous work in 2+1 dimensional quantum gravity, which found evidence for a 
discretization of time in the quantum theory, we reexamine the issue for the case of pure Lorentzian 
gravity with
vanishing cosmological constant and spatially compact universes of genus $g\geq 2$. 
Taking as our starting point the Chern-Simons formulation with
Poincar\'e gauge group, we identify a set of length variables corresponding to space- and timelike 
distances along geodesics in three-dimensional Minkowski space. These are Dirac observables, that is,
functions on the reduced phase space, whose quantization is essentially unique.  
For both space- and timelike distance operators, the spectrum is continuous and not bounded away from
zero.
\end{abstract}

\section{Introduction}

It is not uncommon to hear researchers of quantum gravity express the view that 
spacetime on Planckian distance scales {\it must} possess fundamentally discrete
properties. Given the absence of experimental and observational evidence for or against
such an assertion, and our highly incomplete understanding of quantized gravity,
this points perhaps less to a convergence of different approaches to the problem
of nonperturbative quantum gravity than a shared wish for an ultraviolet cut-off to
render finite certain calculations, for example, of black-hole entropy.\footnote{For a 
reasoning along these lines, see 
\cite{Sorkin:2005qx} and references therein. Related arguments on the existence
of a minimum length scale in quantum gravity can be found in Garay's classic
review \cite{Garay:1994en}.} 
Discussions in the context of popular candidate theories of quantum gravity
in four spacetime dimensions
have revealed numerous subtleties concerning the nature and observability 
of ``fundamental discreteness". Discrete
aspects of asymptotically safe quantum gravity derived from an effective average action
and of loop quantum gravity have been discussed recently in \cite{Reuter:2005bb} and 
\cite{dittrich_arespectra_2007,rovelli_commentarespectra_2007}, respectively. By
contrast, quantum gravity derived from causal dynamical triangulations has so far
not revealed any trace of fundamental discreteness (see, for example, 
\cite{Ambjorn:2008wc}). 
Whether or not Planck-scale discreteness can even in principle be related to 
testable physical phenomena will have to await a deeper understanding of
quantum gravity. 

In this paper, we will address the more specific question of the spectral properties of
quantum operators associated with the length of curves in spacetime, 
and will concentrate on the simpler, non-field 
theoretic setting of pure quantum gravity in 2+1 spacetime dimensions. We will identify 
suitable length functions on the classical reduced phase space and investigate whether
the spectra of their associated quantum operators are continuous or discrete, and whether
this property depends on the time- or spacelike nature of the underlying curves. 
Indications of a possible discrete nature of time in 2+1 dimensional quantum
gravity come from two distinct classical formulations of the theory. 
Firstly, in the so-called polygon approach \cite{hooft_canonical_1993}, based
on piecewise flat Cauchy slicings of spacetime, the Hamiltonian takes the
form of a (compact) angle variable, suggestive of a discrete conjugate time variable in the
quantum theory. Unfortunately, subtleties in the quantization 
\cite{Eldering:2006kk} and the treatment of (residual) gauge 
symmetries \cite{Waelbroeck:1996sg,carlip_quantum_2003} have so far 
prevented a rigorous construction of an
operator implementation of this model. Secondly, an analysis paralleling that of 3+1
loop quantum gravity \cite{freidel_spectra_2003} has also uncovered a discrete spectrum 
for the timelike length operator, albeit at the kinematical level, that is, before imposing 
the quantum Hamiltonian constraint. By contrast, the quantized lengths of spacelike
curves are found to be continuous.

In line with these comments, it should be kept in mind that there is as yet no complete quantization
of three-dimensional Lorentzian gravity for compact spatial slices and in the generic case of
genus $g\geq 2$, which would allow us to settle this question definitively 
(see \cite{Loll:1995kf,Carlip:2004ba} for reviews). There is of course the ``frozen-time" 
Chern-Simons formulation that leads directly to the physical phase space $\mathcal{P}$, the cotangent
bundle of Teichm\"uller space, to which a standard Schr\"odinger quantization can be 
applied \cite{witten_2_1988}. However, as emphasized early on by Moncrief \cite{Moncrief:1990mk},
trying to answer dynamical questions will in general lead to algebraically complicated, 
time-dependent expressions in terms of the canonical variable pairs of this linear phase space, whose 
operator status
in the quantum theory is often unclear. The investigation of time-dependent quantities
is physically meaningful and appropriate, since solutions to the classical Einstein equations in
three dimensions are known to possess initial or final singularities and a nontrivial time development.
As we will see below, this issue is also relevant for the work presented here. 

Beyond the technical problem of identifying well-defined, self-adjoint quantum operators, there
is another layer of difficulty to do with their interpretation and measurability, which is rooted in
the diffeomorphism symmetry of the model, shared with general relativity in four dimensions.
In a gauge theory, physically measurable quantities are usually those which are
invariant under the action of the symmetry group. In the canonical formalism they are also known as 
{\it Dirac observables}. For general relativity they coincide with the diffeomorphism-invariant 
functions on phase space and are necessarily nonlocal \cite{Torre:1993fq}. 
Since time translations form part of the diffeomorphism group, gravitational Dirac observables have the
unusual property of not evolving in time. The usual notion of a time evolution can be recovered
through partially gauge-fixing the diffeomorphism symmetry, a procedure not without its own problems, 
especially when it comes to quantizing the theory. The disappearance of time, and the simultaneous
necessity to select some kind of evolution parameter to describe dynamical processes -- which even
classically is highly non-unique -- form part of the so-called 
\emph{problem of time} in (quantum) gravity \cite{isham_structural_1995}. 
The problems are most severe in the quantum theory, since different
ways of treating time typically give rise to quantum-mechanically inequivalent results, at least in
simple model systems where such results can be obtained explicitly.

Due to their close association with constants of motion, to find gravitational Dirac observables 
one first has to solve the dynamics, at least partially \cite{dittrich_partial_2006}. This is made
difficult by the complexity of the full Einstein equations, and 
hardly any explicit Dirac observables are known\footnote{Fairly general methods for constructing 
Dirac observables have been put forward in 
\cite{rovelli_quantum_2004,dittrich_partial_2006,thiemann_reduced_2006,Pons:2009cz}.}. It is
at this point that our 2+1 dimensional toy model is drastically simpler than the full, 
four-dimensional theory of general relativity:
we can solve its classical dynamics completely and explicitly write down 
the reduced phase space, that is, the space of solutions modulo diffeomorphisms. 
Any function on the reduced phase space corresponds to a Dirac observable and vice versa.

Having a complete set of Dirac observables is not enough; one also needs to know what 
physical observables they represent and -- at least for the case of a realistic theory -- how
they relate to actual, physical measurements. 
Classically, this may not be much of a concern and at most lead to interpretational subtleties,
without affecting calculational results. 
However, during quantization one often has to make a choice of which observables 
are to be represented faithfully as quantum operators, and different choices may well lead to
different conclusions, for example, on the spectral nature of geometric quantum operators.

In this article we study the quantization of a distinguished set of 
geometric observables associated with physical 
lengths and time intervals. Unlike in previous similar investigations, they are genuine Dirac 
observables. 
The quantization of the reduced phase space of our model is straightforward and 
essentially unambiguous, in contrast with the loop quantum gravity approach to 2+1 
dimensional quantum gravity \cite{freidel_spectra_2003}. 
We then present an exact quantization of both space- and timelike length
operators and give a complete analysis of their spectra. 
For the spacelike distances, we find continuous operator spectra, which is perhaps less
surprising. The behaviour of the corresponding operators for timelike distances is 
more subtle. 
It displays certain discrete features, but the length spectrum is not bounded away from zero.
This settles the issue of fundamental discreteness in 2+1 gravity, 
{\it at least for the particular set of
length operators under consideration}, in the negative. Open questions
remain regarding the generality of this result and its relation with physical measurements
in the empty quantum spacetime described by the theory.

The remainder of the paper is organized as follows. 
In the following section 
we review the theory of general relativity in 2+1 dimensions 
with vanishing cosmological constant, and the structure of its reduced phase space with the
standard symplectic structure. We remind the reader that the physical phase space for given 
spacetime topology can be identified with the tangent bundle to a Teichm\"uller space of hyperbolic 
structures on a two-dimensional Riemann surface of genus $g$.
In Sec.\ \ref{sec:observables} we define our distinguished length observables.
The first kind corresponds
to spacelike geodesics in the locally Minkowskian spacetime solutions. In order to obtain also
Dirac observables for timelike lengths, we then define a second kind of variable which 
measures the distance between pairs of such spacelike geodesics. Crucially, we are able
to relate the length variables to well-known functions on Teichm\"uller space. 
We show how the different character of space- and timelike distances in Minkowski space 
translates into
particular angle and length measurements from the viewpoint of hyperbolic geometry.
In Sec.\ \ref{sec:quantization} we quantize both space- and timelike
length observables and analyze their spectra, before presenting our conclusions in 
Sec\ \ref{sec:conclusions}.
In order to make the article more self-contained and some of the derivations
in the main text more explicit, we have collected various mathematical  
results in four appendices. They deal with specific aspects of Lie groups and algebras, 
of hyperbolic geometry and the generalized Weil-Petersson symplectic structure. --
Throughout the article we use units in which $c=16\pi G=1$. In these units the Planck length is 
just equal to $\hbar$.

\section{Gravity in 2+1 dimensions}\label{sec:gravity}

It is well known \cite{wald_general_1984} that a Lorentzian manifold $M$ containing a Cauchy 
surface $\Sigma$ has the product topology $M = \R \times \Sigma$. Moreover, $M$ admits a foliation 
by spacelike surfaces of topology $\Sigma$. In the following we will assume $\Sigma$ to be compact 
and orientable. As a consequence the topology of $\Sigma$, and hence of $M$, is completely characterized 
by the \emph{genus} $g$ of $\Sigma$, the number of holes. 

\subsection{The phase space}\label{sec:phasespace}

Three-dimensional ``general relativity" on the manifold $M$ is defined by the standard Einstein-Hilbert 
action functional of the metric $g$,
\begin{equation} \label{eq:einsteinhilbert}
S[g] = \int_M \rmd^3x \sqrt{-g}\ (R - 2 \Lambda).
\end{equation}
When we take the cosmological constant $\Lambda$ to be zero, the Euler-Lagrange equations have the
familiar form of the vacuum Einstein equations
\begin{equation}
R_{\mu\nu} = 0.
\end{equation}
Gravity in 2+1 dimensions is relatively simple because the Riemann tensor has no additional 
degrees of 
freedom compared to the Ricci tensor \cite{carlip_quantum_2003}, as is clear from the algebraic relation
\begin{equation}
R_{\mu\nu\rho\sigma} = g_{\mu\rho}R_{\nu\sigma} + g_{\nu\sigma}R_{\mu\rho} -g_{\nu\rho}R_{\mu \sigma} - g_{\mu\sigma}R_{\nu\rho}-\frac{1}{2}(g_{\mu\rho}g_{\nu\sigma}-g_{\mu\sigma}g_{\nu\rho})R.
\end{equation}
It follows that solutions to the Einstein equations are flat: any simply connected region in $M$ is isometric 
to a region in three-dimensional Minkowski space. 
The dynamics resides in the transition functions between simply 
connected Minkowski-like regions in a covering of $M$. As we will see below, this information is 
neatly captured by so-called 
holonomies around closed curves in $M$.

Equivalently, we can consider the first-order formulation of the theory. The variables are given by 
two sets of one-forms on $M$, the $\R^3$-valued triad $e^a$ and the $\mathfrak{so}(2,1)$-valued 
spin-connection $\omega^a=\epsilon^{abc}\omega_{bc}$. The Einstein-Hilbert action (\ref{eq:einsteinhilbert}) 
with $\Lambda=0$ now assumes the form
\begin{equation}\label{eq:einsteinhilbertfirstorder}
S[e^a,\omega^a] = -2 \int_M e^a \wedge (\rmd\omega_a + \frac{1}{2} \epsilon_{abc}\omega^b \wedge\omega^c).
\end{equation}
We can combine $e^a$ and $\omega^a$ into a single connection $A$ taking values in the Lie 
algebra $\mathfrak{iso}(2,1)$ of the Poincar\'e group (see \ref{sec:gaugegroup}). 
In terms of the Poincar\'e-connection $A$ the action (\ref{eq:einsteinhilbertfirstorder}) 
up to boundary terms takes the form of a Chern-Simons action 
\cite{witten_2_1988, carlip_quantum_2003}, namely,
\begin{eqnarray}
S[A] &= - \int_M \Tr_B (A \wedge \rmd A + \frac{2}{3} A\wedge A\wedge A)\nonumber\\
& = - \int_M \rmd x\,\epsilon^{\lambda\mu\nu}B\left( A_{\lambda},(\rmd A)_{\mu\nu} + \frac{2}{3}\left[A_{\mu},A_{\nu}\right]\right)\label{eq:chernsimons},
\end{eqnarray}
where $B$ is the bilinear form on $\mathfrak{iso}(2,1)$ defined in \ref{sec:gaugegroup}.
Denoting the curvature of $A$ by $F(A)=\rmd A + A\wedge A$, the equations of motion are simply given
by $F(A)=0$. 

The \emph{Poincar\'e holonomy} along a closed curve $\gamma$ in $M$ based at a point $x_0$ 
(together with a chosen basis of the tangent space at $x_0$) is defined as the path-ordered 
exponential
\begin{equation}\label{eq:poincareholonomy}
g_{\gamma,x_0} = \mathcal{P} \exp \int_{\gamma} A\ \in ISO(2,1)
\end{equation}
taking values in the Poincar\'e group.
The vanishing curvature of $A$ implies that $g_{\gamma,x_0}$ is invariant under deformations of $\gamma$,
up to conjugation. As a consequence, for a given connection $A$ the holonomy is only a function 
of the 
homotopy class $[\gamma]$ of the closed curve $\gamma$. Solutions to the equations of motion 
are characterized by their holonomies. 
More precisely, Mess \cite{mess_lorentz_2007} (see also \cite{andersson_notespaper_2007}) 
has proved that any suitable homomorphism from the fundamental group $\pi_1$ to $ISO(2,1)$ 
corresponds to a unique maximal flat spacetime\footnote{For a {\it maximal flat spacetime} $M$ 
any isometric imbedding in a flat spacetime N is necessarily surjective.}, leading to
the identification
\begin{equation}\label{eq:phasespaceiso}
\mathcal{P} = \mathrm{Hom}_0(\pi_1,ISO(2,1))/ISO(2,1),
\end{equation}
for the phase space $\mathcal{P}$. The subscript ``0" indicates a restriction to those 
homomorphisms whose $SO(2,1)$-projections have a Fuchsian subgroup of $SO(2,1)$ as
image (see \ref{sec:hyperbolic geometry}). Note that the fundamental group $\pi_1$ of $M$
is equal to the fundamental group of the spacelike surface $\Sigma$.

Now that we have learned how to assign a set of Poincar\'e holonomies to a flat spacetime,
can we also achieve the converse, that is, reconstruct the flat spacetime (by identifying
points in Minkowski space) from a given homomorphism $\phi: \pi_1 \rightarrow ISO(2,1)$?
It was proved in \cite{mess_lorentz_2007} that there exists a unique convex open subset $U$ of Minkowski space on which $\phi$ acts properly discontinuously, giving rise to a quotient space 
of $U$ which is a maximal spacetime, necessarily having the right holonomies. 
Constructing the subset $U$ is difficult for general $\phi$, but can be obtained in a 
constructive way for a dense subset of phase space by the method of \emph{grafting} \cite{mcmullen_complex_1998,benedetti_cosmological_2001,meusburger_grafting_2006}.
 
A space closely related to the phase space $\mathcal{P}$ is \emph{Teichm\"uller space}
\begin{equation}
\mathcal{T}=\mathrm{Hom}_0(\pi_1(\Sigma),PSL(2,\R))/PSL(2,\R),
\end{equation}
describing the space of conformal or complex structures on the surface $\Sigma$ 
(\ref{sec:hyperbolic geometry}). Identifying $PSL(2,\R)$ with the future-preserving Lorentz group 
$SO_0(2,1)$ (\ref{sec:gaugegroup}), it is immediately clear that we obtain a canonical projection 
$\pi_{\mathcal{T}}$ of $\mathcal{P}$ onto $\mathcal{T}$ by simply taking the 
$SO_0(2,1)\cong PSL(2,\R)$-part of the $ISO_0(2,1)$-holonomies. It turns out that 
$\pi_{\mathcal{T}}$ identifies $\mathcal{P}$ with the tangent bundle of Teichm\"uller space: 
given a path $t\rightarrow[\phi](t)$ in $\mathcal{T}$, first taking the derivative with respect to $t$ and evaluating on a homotopy class, and then reversing the order gives a correspondence
\begin{eqnarray}
T\mathcal{T} &=&T(\mathrm{Hom}_0(\pi_1(\Sigma),PSL(2,\R))/PSL(2,\R))\nonumber\\ 
& \cong & \mathrm{Hom}_0(\pi_1(\Sigma),TPSL(2,\R))/TPSL(2,\R).
\end{eqnarray}
Using the fact that $T\,PSL(2,\R)$ and $ISO_0(2,1)$ are isomorphic (\ref{sec:gaugegroup}), 
we conclude that
\begin{equation}\label{eq:phasespacetangentbundle}
\mathcal{P}=T\,\mathcal{T}.
\end{equation}

\subsection{Symplectic structure}\label{sec:symplecticstructure}

To obtain the symplectic structure on $\mathcal{P}$ we foliate
(\ref{eq:einsteinhilbertfirstorder}) into constant-time slices and identify the 
canonical momenta, leading to the basic Poisson brackets
\begin{equation}
\{e^a_i(x),\omega^b_j(y)\} = -\frac{1}{2}\epsilon_{ij}\eta^{ab}\delta(x,y),
\end{equation}
where the one-forms $e^a$ and $\omega^a$ are restricted to a constant-time surface $\Sigma$. 
In terms of the connection $A$ we can write the symplectic structure as the two-form
\begin{equation}
\Omega = \int_{\Sigma} \Tr( \delta A\wedge\delta A )
\end{equation}
on the (infinite-dimensional) space of connections restricted to $\Sigma$,
which descends to a symplectic structure $\omega$ on the space $\mathcal{P}$ of flat connections. 
It can be shown \cite{alekseev_symplectic_1995,atiyah_yang-mills_1983} that for connections in 
a general gauge group $G$ this $\omega$ corresponds to a canonical symplectic 
structure \cite{goldman_invariant_1986} on $\mathrm{Hom}_0(\pi_1(\Sigma),G)/G$ which 
is a generalization of the well-known Weil-Petersson symplectic structure $\omega_{WP}$ 
in the case of $G = PSL(2,\R)$ (see \ref{sec:hyperbolic geometry}).

We expect this \emph{generalized Weil-Petersson symplectic structure} (\ref{sec:weilpetersson})
corresponding to the tangent group $TPSL(2,\R)$ to be related to the standard Weil-Petersson 
structure $\omega_{WP}$ on $\mathcal{T}$. Indeed, it is straightforward to associate to the 
tangent bundle of a symplectic manifold a canonical symplectic structure, the 
\emph{tangent symplectic structure} \cite{grabowski_tangent_2007}. To see this, note that 
the two-form $\omega_{WP}$ defines a linear map 
\begin{equation}
\tilde{\omega}_{WP} : T\,\mathcal{T} \to T^*\,\mathcal{T}
\end{equation}
by contraction. At the same time, the cotangent bundle $T^*\,\mathcal{T}$ already 
possesses a canonical symplectic structure $\omega_{can}$, which we can pull back along 
$\tilde{\omega}_{WP}$ to obtain a symplectic form
\begin{equation}
\omega = \tilde{\omega}_{WP}^*\,\omega_{can}
\end{equation}
on $T\,\mathcal{T}$. We show in \ref{sec:weilpetersson} that this coincides with the generalized 
Weil-Petersson symplectic structure for the tangent group.

The relation between $\omega$ and $\omega_{WP}$ is most transparent when we look at 
the Poisson brackets they define. Given a function $f$ on $\mathcal{T}$, 
there are {\it two} functions on $\mathcal{P}=T\,\mathcal{T}$ we can naturally associate with it.
First, we can just take the trivial extension $f \circ \pi_{\mathcal{T}}$ of $f$, which 
we will continue to denote by $f$. Second, we can take the derivative $\rmd f:T\,\mathcal{T}\to\R$, 
which we call the \emph{variation} of $f$, and which in the following we will often denote 
by the corresponding capital letter $F$. 
The relation between the two different Poisson brackets can be summarized by \cite{grabowski_tangent_2007}
\begin{eqnarray}\label{eq:tangentpoissonbracket}
\{ f_1, f_2\}_{\mathcal{P}} = 0, \nonumber\\
\{ \rmd f_1, f_2\}_{\mathcal{P}}  =  \{f_1,f_2\}_{\mathcal{T}}, \\
\{ \rmd f_1,\rmd f_2 \}_{\mathcal{P}}  =  \rmd \{f_1,f_2\}_{\mathcal{T}}\nonumber
\end{eqnarray}
for any pair $f_1$ and $f_2$ of functions on Teichm\"uller space.

Let us check explicitly that (\ref{eq:tangentpoissonbracket}) yields the Poisson brackets familiar 
from the literature. Following \cite{carlip_quantum_2003}, define the loop variable 
$T^0[\gamma] := \frac{1}{2}\Tr g_{\gamma}$, where $g_{\gamma}$ is the $SO(2,1)$-holonomy 
around $\gamma$, analogous to (\ref{eq:poincareholonomy}) above, and 
its variation by $T^1[\gamma]:=\rmd T^0[\gamma]$. 
For their Poisson brackets, we derive \cite{carlip_quantum_2003} 
\begin{eqnarray}
\left\{T^0[\gamma_1],T^0[\gamma_2]\right\} = 0 \nonumber\\
\left\{T^1[\gamma_1],T^0[\gamma_2]\right\} = -\frac{1}{2} \sum_i \epsilon(p_i) 
\left(T^0[\gamma_1 \circ_i \gamma_2] - T^0[\gamma_1 \circ_i \gamma_2^{-1}]\right) 
\label{eq:T0T1fromcarlip}\\
\left\{T^1[\gamma_1],T^1[\gamma_2]\right\} = -\frac{1}{2} \sum_i \epsilon(p_i) 
\left(T^1[\gamma_1 \circ_i \gamma_2] - T^1[\gamma_1 \circ_i \gamma_2^{-1}]\right),
\nonumber
\end{eqnarray}
where $\gamma_1 \circ_i \gamma_2$ denotes the path obtained by cutting open
$\gamma_1$ and $\gamma_2$ at the $i$'th intersection point $p_i$ and composing them
with the curve orientations as indicated, and $\epsilon(p_i)=\pm 1$ depending on the relative
orientation of the two tangent vectors.
Clearly, (\ref{eq:T0T1fromcarlip}) is of the form of (\ref{eq:tangentpoissonbracket}) 
if the Poisson bracket {\it on Teichm\"uller space} is given by
\begin{equation}
\left\{T^0[\gamma_1],T^0[\gamma_2]\right\}_{\mathcal{T}} = -\frac{1}{2} \sum_i \epsilon(p_i) \left(T^0[\gamma_1 \circ_i \gamma_2] - T^0[\gamma_1 \circ_i \gamma_2^{-1}]\right).
\end{equation}
However, according to \cite{goldman_invariant_1986} this is precisely the Poisson bracket 
we get for the generalized Weil-Petersson structure for the group $SO(2,1)$. 
Due to the isomorphism between $SO(2,1)$ and $PSL(2,\R)$ it corresponds to the standard 
Weil-Petersson symplectic structure on Teichm\"uller space. Finally, note that by construction 
the map $\tilde{\omega}_{WP}$ is an isomorphism of symplectic manifolds which identifies 
the phase space $\mathcal{P}$ with the cotangent bundle of Teichm\"uller space. 
This will make the quantization of the theory in Sec.\ \ref{sec:quantization} straightforward.

\section{Geometric observables}\label{sec:observables}
 
In the previous section we have established a full correspondence between the phase 
space $\mathcal{P}$ and the tangent bundle to Teichm\"uller space. The latter is well
studied and has a nice description in terms of hyperbolic geometry on Riemann surfaces 
(see \ref{sec:hyperbolic geometry}). We will now show how particular observables in our 2+1 dimensional spacetime can be interpreted as variations of geometric functions on 
Teichm\"uller space. 

\begin{figure}
\begin{center}
\includegraphics[width=10cm]{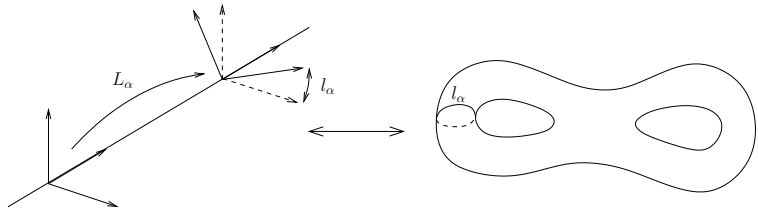}
\caption{A Poincar\'e transformation $(g_\alpha,X_\alpha)$ leaving a geodesic invariant can be
described as a translation by $L_\alpha$ along the geodesic, followed by a boost of
rapidity $l_\alpha$ in the plane perpendicular to the geodesic (left). At the same time, 
$l_\alpha$ can be thought of as the length of a unique closed geodesic on an associated
Riemann surface (right).
\label{figure:invariantgeodesic}}
\end{center}
\end{figure}

Let us first examine how a Poincar\'e holonomy acts on Minkowski space. 
We will restrict ourselves to transformations which describe boosts, since the Lorentzian parts 
of the nontrivial holonomies in (\ref{eq:phasespaceiso}) are necessarily hyperbolic 
(\ref{sec:hyperbolic geometry}). In the following we will often identify Minkowski space with the Lie algebra $\mathfrak{sl}(2,\R)$ together with its indefinite metric $B$ (as spelled out in 
\ref{sec:gaugegroup}), and $ISO(2,1)$ with $PSL(2,\R)\ltimes\mathfrak{sl}(2,\R)$. 
A holonomy $(g,X)\in PSL(2,\R)\ltimes\mathfrak{sl}(2,\R)$ then acts on Minkowski space by
\begin{equation}
Y \to \Ad(g)Y + X.
\end{equation}
If $g$ is nontrivial, $\Ad(g)$ will leave exactly one direction invariant, which according to 
(\ref{eq:adginvariant}) is given by $\xi_l(g)$.\footnote{Here, $\xi_l(g)$ is the variation of
the hyperbolic length function $l(g)$ on $PSL(2,\R)$ defined in (\ref{eq:explgxilg}).} 
For $(g,X)$ to leave a geodesic in Minkowski space invariant, the latter must be aligned 
with the invariant direction $\xi_l(g)$, and thus  can be parametrized as 
$t\to Y+t\,\xi_l(g)$. It is invariant if and only if
\begin{equation}\label{eq:invariantgeodesic}
\Ad(g)(Y+t \,\xi_l(g))+X = Y + (t+L)\xi_l(g)
\end{equation}
for some $L\in\R$ and all $t$. If we denote by $P_{\perp}$ the projection onto the subspace 
$\xi_l(g)_{\perp}\subset\mathfrak{sl}(2,\R)$ perpendicular to $\xi_l(g)$, it follows that we must 
have
\begin{equation}\label{eq:perpequationforY}
(\Ad(g)-1)P_{\perp}(Y) = -P_{\perp}(X).
\end{equation}
Since $\Ad(g)-1$ is a bijection when restricted to $\xi_l(g)_{\perp}$, eq.\ (\ref{eq:perpequationforY}) 
has a unique solution for $Y$ up to a shift in the direction $\xi_l(g)$. It is not hard to see that this 
solves (\ref{eq:invariantgeodesic}) when we take $L$ to be
\begin{equation}
L = B(\xi_l(g),X).
\end{equation}
Note that by construction $\xi_l(g)$ is spacelike and of unit norm (c.f. \ref{sec:grouptheory}), 
which implies that $L$ is a {\it space}like distance. We conclude that we can describe a 
hyperbolic Poincar\'e transformation as a translation by a distance $L$ along a geodesic 
followed by a boost in the plane perpendicular to the geodesic (see the left-hand side of 
Fig.\ \ref{figure:invariantgeodesic}). From (\ref{eq:explicitexpad}) it follows that
\begin{equation}\label{eq:dotproductboost}
B(Z,\Ad(g)Z)=\cosh l(g)
\end{equation}
for a unit vector $Z$ perpendicular to $\xi_l(g)$. We deduce that $l(g)$ is precisely the boost parameter 
(or change of rapidity). 

Suppose now we are given a spacetime solution $M\in\mathcal{P}$. 
For any closed curve $\alpha$ in $M$ we get a Poincar\'e holonomy $(g_{\alpha},X_{\alpha})$
and two associated phase space functions 
\begin{eqnarray}
l_{\alpha} &=& l(g_{\alpha}),\\
L_{\alpha} &=& B(\xi_l(g_{\alpha}),X_{\alpha}).
\end{eqnarray}
From the definition (\ref{eq:xidef}) it is clear that $L_{\alpha}$ is just the variation of $l_{\alpha}$,
\begin{equation}\label{eq:lengthvariation}
L_{\alpha} = \rmd l_{\alpha} : T\,\mathcal{T}\to \R.
\end{equation}
What is the interpretation of the observable $L_{\alpha}$? As we have mentioned earlier, 
the spacetime $M$ can be reconstructed by taking the quotient of a subset $U$ of Minkowski 
space by the action of all holonomies. 
Thus, \emph{if} the geodesic invariant under $(g_{\alpha},X_{\alpha})$ would lie inside $U$, 
it would descend to a closed geodesic of length $L_{\alpha}$ in $M$ homotopic to $\alpha$. 
Moreover, it would be the path with minimal length in the homotopy class. Unfortunately $U$ 
is necessarily a convex subset and therefore cannot contain any complete geodesic. 
This means that when we try to minimize the length of a path in a homotopy class, we will
necessarily run into the initial singularity of the spacetime. 
We will nevertheless work with $L_{\alpha}$ as a 
geometric observable which probes the length scales of the spacetime manifold and  
somewhat inaccurately refer to it as the ``length of the closed geodesic $\alpha$ in $M$". We will
return to this issue in the discussion section.

The function $L_{\alpha}$ has already been studied in a slightly different form in the mathematics 
literature, where it is referred to as the \emph{Margulis invariant} \cite{goldman_margulis_2002}. 
In the work of Meusburger \cite{meusburger_cosmological_2008} $l_{\alpha}$ and 
$L_{\alpha}$ are called the \emph{mass} and \emph{spin} of $\alpha$ and are used as a 
complete set of observables on phase space. In terms of hyperbolic geometry 
(\ref{sec:hyperbolic geometry}) the function $l(g_{\alpha})$ can be interpreted 
as the hyperbolic length of the unique closed geodesic homotopic to 
$\alpha$ on the Riemann surface (Fig.\ \ref{figure:invariantgeodesic}, right). 

Since the lengths $L_{\alpha}$ only probe spacelike distances, we will now define a new observable, 
the distance between two closed geodesics, which can be either space- or timelike. 
Let $\alpha_1$, $\alpha_2$ be two closed paths in $M$ and denote their holonomies by 
$(g_1,X_1),(g_2,X_2)\in PSL(2,\R) \ltimes \mathfrak{sl}(2,\R)$,
with $\gamma_1$, $\gamma_2$ the associated invariant geodesics in Minkowski space. 
We are interested in the line-segment $c$ connecting $\gamma_1$ and $\gamma_2$
at right angles. Since the directions of the geodesics are given by
$\xi_l(g_1)$ and $\xi_l(g_2)$, the direction of $c$ will be their cross product,
which in Lie algebra terms is just the commutator $[\xi_l(g_1),\xi_l(g_2)]$. For two
points $Y_1, Y_2\in\mathfrak{sl}(2,\R)$ on the two geodesics $\gamma_1$ and $\gamma_2$, 
the signed length of $c$ is equal to
\begin{equation}\label{eq:definitiondistance}
D_{\alpha_1\alpha_2}= \frac{B\left(Y_1-Y_2,[\xi_l(g_1),\xi_l(g_2)]\right)
}{\sqrt{\left|B([\xi_l(g_1),\xi_l(g_2)],[\xi_l(g_1),\xi_l(g_2)])\right|}}.
\end{equation}
For 3-vectors $x^a$ and $y^a$ we have the identity
\begin{equation}
(x\times y)\cdot(x\times y) = x^ay^b\epsilon_{ab}^{\,\,c}x^{a'}y^{b'}\epsilon_{a'b'}^{\,\,c'}\eta_{cc'} = (x^ay_a)^2-(x^ax_b)(y^by_b),
\end{equation}
which in our case implies
\begin{equation}
B([\xi_l(g_1),\xi_l(g_2)],[\xi_l(g_1),\xi_l(g_2)])=B(\xi_l(g_1),\xi_l(g_2))^2-1,
\end{equation}
where we have used that $\xi_l(g_i)$ is of unit norm. Consequently, $[\xi_l(g_1),\xi_l(g_2)]$ 
is spacelike when $B(\xi_l(g_1),\xi_l(g_2))>1$ and timelike when $B(\xi_l(g_1),\xi_l(g_2))<1$.

\begin{figure}
\begin{center}
\includegraphics[width=6cm]{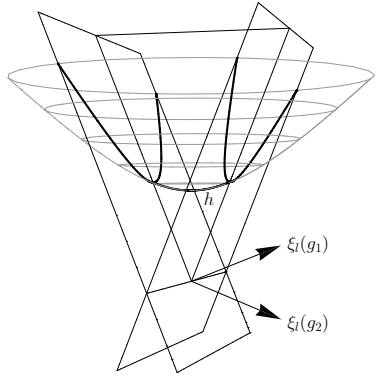}
\caption{The distance $h$ between two geodesics in the unit hyperboloid $H_1$ is the 
hyperbolic angle between 
the corresponding planes through the origin in Minkowski space.\label{figure:distanceinhyperboloid}}
\end{center}
\end{figure}

\begin{figure}
\begin{center}
(a) \includegraphics[width=11cm]{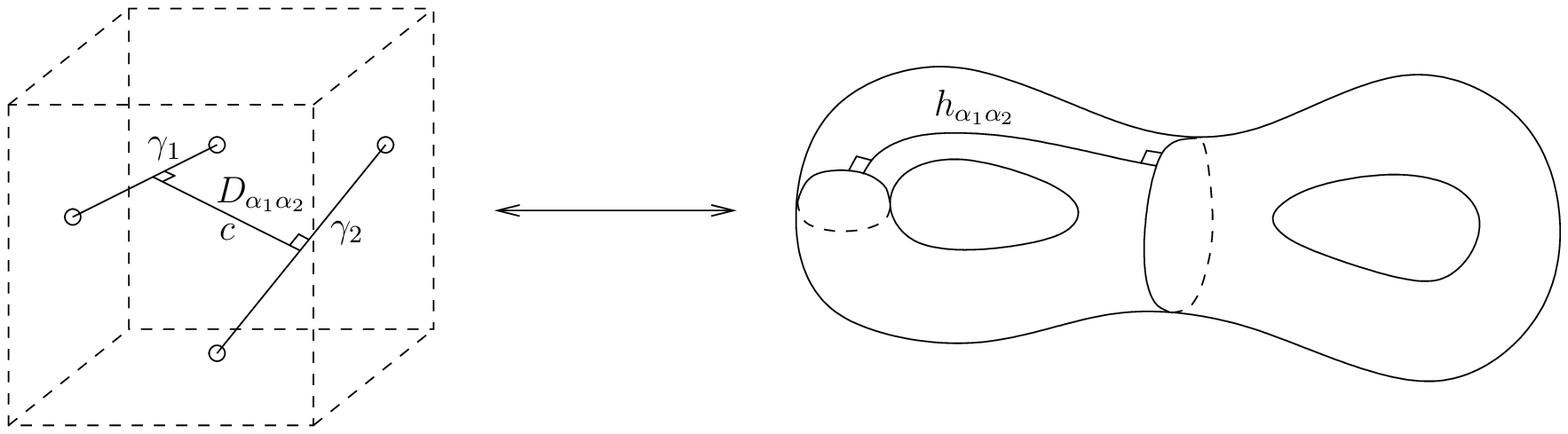} \\
(b) \includegraphics[width=11cm]{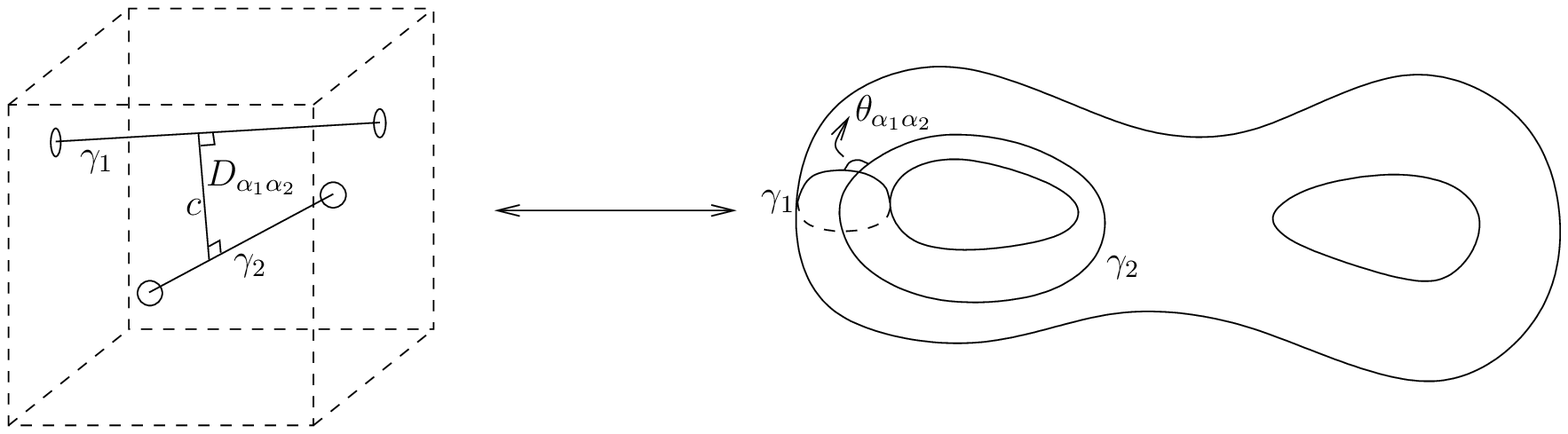}
\caption{Distance $D_{\alpha_1\alpha_2}$ between (a) spacelike and (b)
timelike separated closed geodesics in Minkowski space and their relation to hyperbolic
geometry. \label{figure:geodesicdistance}}
\end{center}
\end{figure}

This raises the interesting question of how the two cases differ at the level of hyperbolic geometry. 
It turns out that when the two closed geodesics on the Riemann surface are non-intersecting 
(Fig.\ \ref{figure:geodesicdistance}a), we have
\begin{equation}\label{eq:xiproductdistance}
|B(\xi_l(g_1),\xi_l(g_2))|=\cosh h > 1,
\end{equation}
where $h$ is the (shortest) hyperbolic distance between the two. If they do intersect (Fig.\ \ref{figure:geodesicdistance}b), we have
\begin{equation}\label{eq:xiproductangle}
B(\xi_l(g_1),\xi_l(g_2)) = \cos \theta < 1,
\end{equation}
where $\theta$ is the angle between the geodesics at the intersection point. 
A simple way to see this is by considering the hyperboloid model $H_1$ (as described in
\ref{sec:hyperbolic geometry}). 
The two geodesics define two planes through the origin in Minkowski space with normals 
equal to $\xi_l(g_1)$ and $\xi_l(g_2)$ (Fig.\ \ref{figure:distanceinhyperboloid}), 
and intersection spanned by the outer product $[\xi_l(g_1),\xi_l(g_2)]$. 
The intersection will obviously only intersect $H_1$ if the two geodesics intersect on $H_1$, 
therefore $[\xi_l(g_1),\xi_l(g_2)]$ is timelike if and only if the two geodesics intersect. 
Now there is a unique element $g\in PSL(2,\R)$ for which $\Ad(g)$ maps $\xi_l(g_1)$ to 
$\xi_l(g_2)$ and leaves $[\xi_l(g_1),\xi_l(g_2)]$ invariant. 

If the commutator $[\xi_l(g_1),\xi_l(g_2)]$ is spacelike, the group element $g$ is hyperbolic and 
$\xi_l(g)$ is proportional to $[\xi_l(g_1),\xi_l(g_2)]$. From (\ref{eq:dotproductboost}) we deduce
that the scalar product of the two vectors is given by $B(\xi_l(g_1),\xi_l(g_2))=\cosh l(g)$. 
The invariant geodesic in $H_1$ corresponding to $g$ 
is the intersection of the plane spanned by $\xi_l(g_1)$ and $\xi_l(g_2)$ with $H_1$. 
It therefore coincides with the perpendicularly connecting geodesic, and the searched-for distance 
$h$ is just $l(g)$. On the other hand,
if $[\xi_l(g_1),\xi_l(g_2)]$ is timelike, the angle $\theta$ between the geodesics in $H_1$ 
is just the angle between the two planes, which satisfies $B(\xi_l(g_1),\xi_l(g_2))=\cos \theta$.

In order to calculate the variation of $B(\xi_l(g_1),\xi_l(g_2))$ (that is, of $h$ and $\theta$) to arrive
at the Dirac length observable, we first need an identity for the derivative of $\xi_l(g)$, namely,
\begin{equation}
\left.\frac{\rmd}{\rmd t}\right|_{t=0} \xi_l(\exp(tX)g) = [Y,\xi_l(g)],
\end{equation}
where $Y$ is a point on the invariant geodesic. 
To prove this, note that $\xi_l(\exp(t \xi_l(g))g)=\xi_l(g)$ for all $t$, which means that we can 
replace $X$ by $(\Ad(g)-1)Y$ according to (\ref{eq:perpequationforY}), 
\begin{eqnarray}
\left.\frac{\rmd }{\rmd t}\right|_{t=0} \xi_l(\exp(tX)g) &=& \left.\frac{\rmd }{\rmd t}\right|_{t=0}\xi_l(\exp(t(\Ad(g)-1)Y)g )\nonumber\\
&=&\left.\frac{\rmd }{\rmd t}\right|_{t=0}\xi_l(\exp(t Y)g\exp(-t Y) )\nonumber\\
&=&\left.\frac{\rmd }{\rmd t}\right|_{t=0}\Ad(\exp(t Y))\xi_l(g)\nonumber\\
&=&\left.\frac{\rmd }{\rmd t}\right|_{t=0}\exp(t\ \ad(Y))\xi_l(g)\nonumber\\
&=& \ad(Y)\xi_l(g) = [Y,\xi_l(g)].
\end{eqnarray}
Using this result, we find for the variation of $B(\xi_l(g_1),\xi_l(g_2))$
\begin{eqnarray}
\left.\frac{\rmd }{\rmd t}\right|_{t=0}B(\xi_l(\exp(t X_1)g_1),\xi_l(\exp(t X_2)g_2) ) \nonumber\\ 
= B\left(\left.\frac{\rmd }{\rmd t}\right|_{t=0}\xi_l(\exp(t X_1)g_1),\xi_l(g_2)\right)\nonumber\\
\quad + B\left(\xi_l(g_1),\left.\frac{\rmd }{\rmd t}\right|_{t=0}\xi_l(\exp(t X_2)g_2)\right) \nonumber\\
=B([Y_1,\xi_l(g_1)],\xi_l(g_2))+B(\xi_l(g_1),[Y_2,\xi_l(g_2)]) \nonumber\\
=B(Y_1-Y_2,[\xi_l(g_1),\xi_l(g_2)]),
\end{eqnarray}
so that finally
\begin{equation}
\rmd h = \frac{\rmd \cosh h}{\sqrt{\cosh^2h-1}} = \frac{B(Y_1-Y_2,[\xi_l(g_1),\xi_l(g_2)])}{\sqrt{B(\xi_l(g_1),\xi_l(g_2))^2-1}},
\end{equation}
and similarly for $\theta$. We conclude that
\begin{equation}
D_{\alpha_1\alpha_2} = \left\{\begin{array}{cl}\rmd \theta_{\alpha_1\alpha_2} & 
\alpha_1,\alpha_2\;\mathrm{intersect\ on\ Riemann\ surface} \\ \rmd h_{\alpha_1\alpha_2} & \mathrm{otherwise}\end{array}\right.,
\end{equation}
which is illustrated in Fig.\ \ref{figure:geodesicdistance}.

From relations (\ref{eq:xiproductdistance}) and (\ref{eq:xiproductangle}) it is now straightforward 
to give a geometric interpretation of the functions $\theta_{\alpha_1\alpha_2}$ and $h_{\alpha_1\alpha_2}$ on $\mathcal{P}$: 
$h$ is the hyperbolic angle (or boost parameter) between $\gamma_1$ and $\gamma_2$ measured along 
the connecting geodesic $c$, and $\theta$ is the angle between $\gamma_1$ and $\gamma_2$ along $c$.
With regard to our quest for expressing geometric quantities in Minkowski space in terms of
``Teichm\"uller data", we can already see the general picture emerging: spacelike geodesics in Minkowski 
space are related to geodesics on the Riemann surface, and distances along them in Minkowski space 
correspond to variations of hyperbolic distances. By contrast, timelike geodesics relate to points in the 
Riemann surface, and timelike distances correspond to variations of angles at those points.

\section{Quantization}\label{sec:quantization}

In section \ref{sec:symplecticstructure} we identified the phase space $\mathcal{P}$ of 2+1 
dimensional gravity with the cotangent bundle $T^*\mathcal{T}$ to Teichm\"uller space together
with its canonical symplectic structure.
Geometric quantization of this phase space is straightforward. As Hilbert space we take  $\mathcal{H}=L^2(\mathcal{T},\omega_{WP}^{3g-3})$, the space of square-integrable wave
functions on Teichm\"uller space with volume form defined by the Weil-Petersson symplectic 
structure $\omega_{WP}$. A function $f$ on Teichm\"uller space becomes a multiplication operator
\begin{equation}
\hat{f}\phi = f\cdot\phi,
\end{equation}
and its variation $F$ a derivative operator according to
\begin{equation}
\hat{F}\phi = \rmi \hbar \{f,\phi\}_{WP}.
\end{equation}
One easily checks that this yields an operator representation of the Poisson algebra 
(\ref{eq:tangentpoissonbracket}) of phase space functions at most linear in the translational part 
of the holonomies. By the Stone-von Neumann theorem the quantization of the latter algebra is 
unique up to unitary equivalence, because our phase space can be brought globally to the 
canonical form $T^* \R^{6g-6}$.\footnote{By contrast, the so-called moduli space 
$\mathcal{M}:=
\mathcal{T}/\mathcal{MCG}$, obtained by taking a quotient with respect to the mapping class
group $\mathcal{MCG}$ of ``large diffeomorphisms" (generated by Dehn twists), 
is not simply connected. 
Some of the difficulties which arise when implementing $\mathcal{MCG}$ as a symmetry group
either in the classical or the quantum theory are described in \cite{carlip_quantum_2003}.}

The procedure for finding the spectrum of an operator $\hat{F}$ corresponding to the variation 
of a function $f$ on Teichm\"uller space $\mathcal{T}$ is relatively straightforward. 
The Hamiltonian 
vector field $H_f = \tilde{\omega}_{WP}^{-1}(\rmd f)$ generates the Hamiltonian flow of $f$ on 
$\mathcal{T}$. If we take a wave function $\phi$ with support on a single orbit 
$\mathcal{O}$ of the flow, it will be an eigenstate of $\hat{F}$ with eigenvalue $F$ if it 
describes a wave in the flow parameter $t$, that is,
\begin{equation}
\phi|_{\mathcal{O}}(t) \propto \exp(-\frac{\rmi}{\hbar}F t).
\end{equation}
Whether the spectrum of $\hat{F}$ (restricted to the orbit $\mathcal{O}$) is continuous or discrete
depends on the domain of $t$. Whenever the flow is well-defined and injective for $t\in\R$, 
$F$ can take any value in $\R$. However, if $t$ is restricted to take values in a bounded interval, 
say, $t\in ]0,r[$, we can only have a discrete set of eigenstates with eigenvalues $F$ which 
are separated by a distance $2\pi\hbar/r$. The precise eigenvalues depend on the chosen 
self-adjoint extension of $\hat{F}$ or, equivalently, on the chosen boundary conditions for 
$\phi$. To get the full spectrum of $F$ we must combine all spectra of the individual orbits, 
which need not coincide.

\subsection{Spectra of length observables}

\begin{figure}
\begin{center}
\centering
\includegraphics[width=5cm]{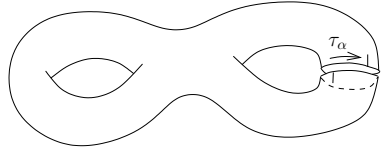}
\caption{The Hamiltonian vector field $H_{l_{\alpha}}$ generates a twist along $\alpha$. \label{figure:twist}}
\end{center}
\end{figure}

Recall that the length of a closed geodesic is given by the variation $L_{\alpha}=\rmd l_{\alpha}$ 
of the hyperbolic length $l_{\alpha}$. A convenient global coordinate system for Teichm\"uller
space is given by the 
Fenchel-Nielsen coordinates $(l_i,\tau_i)$, $i=1,\dots,3g-3$ (see \ref{sec:hyperbolic geometry}), 
corresponding to a pair-of-pants 
decomposition which has $\alpha$ as one of the cuts. In these coordinates the Weil-Petersson symplectic 
form is given by (\ref{eq:weilpeterssoninfenchelnielsen}),
\begin{equation}
\omega_{WP} = \sum_i \rmd l_i\wedge \rmd \tau_i,
\end{equation}
where the $\tau_i$ are the twist parameters. The Hamiltonian flow of $l_{\alpha}$ is simply the 
twist flow along $\alpha$ (Fig.\ \ref{figure:twist}). Since the twist parameters as coordinates 
on Teichm\"uller space have domain equal to $\R$, we conclude that the spectrum of 
$\hat{L}_{\alpha}$ is the entire real line $\R$. 

\begin{figure}
\begin{center}
\centering
\begin{tabular}{cc}
(a) \includegraphics[width=5cm]{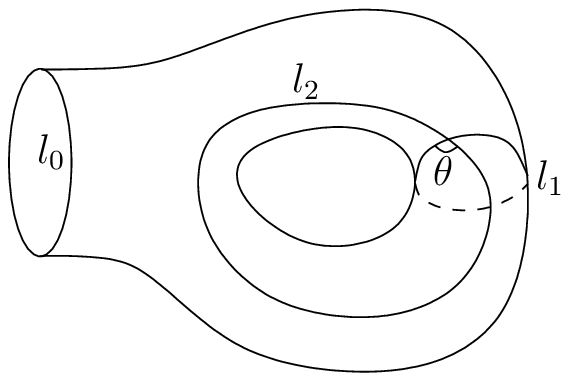} & (b)
\includegraphics[width=5cm]{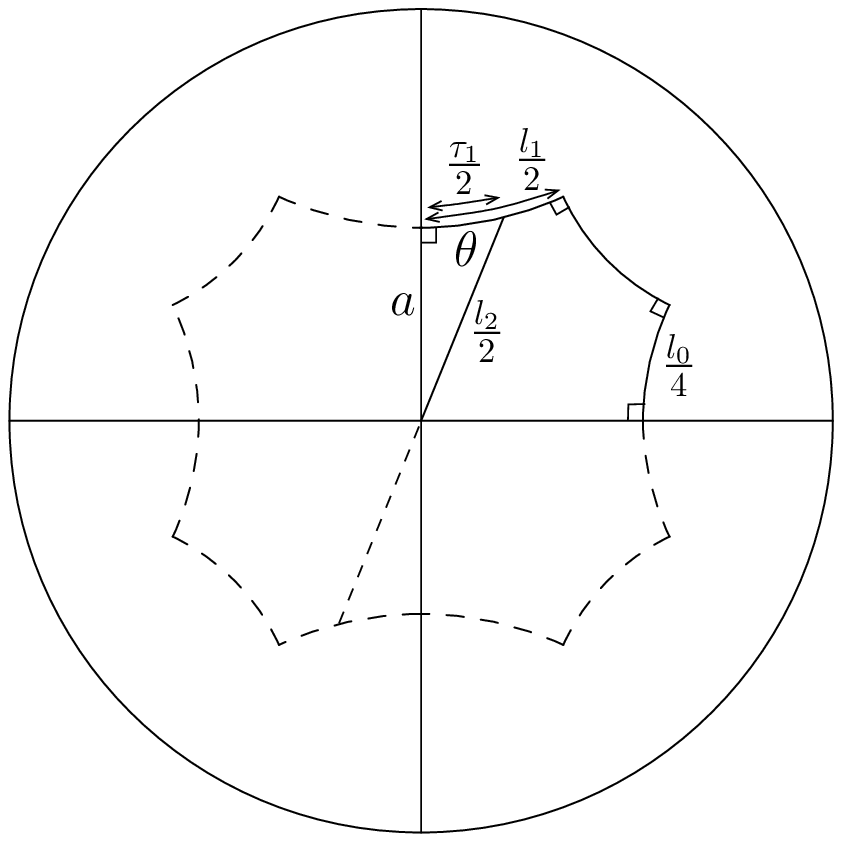} 
\end{tabular}
\caption{(a) Angle between two geodesics on the one-holed torus. (b) Cutting the torus open,
one obtains an octagon, which can be put inside the Poincar\'e disc.
\label{figure:angleoneholedtorus}}
\end{center}
\end{figure}
Next, we will investigate the operator 
\begin{equation}\label{dop}
\hat{D}_{\alpha_1\alpha_2}=\rmi \hbar\{\theta,\cdot\}_{WP}
\end{equation}
corresponding to a timelike distance between two geodesics. To start with, consider the geometric
situation as depicted in Fig.\ \ref{figure:angleoneholedtorus}, namely, a Riemann surface of genus 1 
with a hole of geodesic boundary length $l_0$. Fixing $l_0$ means that its hyperbolic geometry 
is described by a two-dimensional Teichm\"uller space $\mathcal{T}$. 
Once we have found the spectrum of $\hat{D}_{\alpha_1\alpha_2}$, we will argue that the result 
holds for any spatial topology 
and for any two simple closed geodesics $\alpha_1$ and $\alpha_2$ with a single intersection. 

One way of parametrizing $\mathcal{T}$ (up to a sign) is through the lengths $l_1$ and $l_2$ of 
$\gamma_1$ and $\gamma_2$ as indicated in Fig.\ 
\ref{figure:angleoneholedtorus}. Cutting the surface along $\gamma_1$ and along two shortest
geodesics connecting the hole's boundary to either side of $\gamma_1$, we obtain an 
eight-sided polygon which we can draw at the centre of the 
Poincar\'e disc as in Fig.\ \ref{figure:angleoneholedtorus}. Using the symmetries of the situation
and applying the trigonometric 
identities from \ref{sec:hyperbolic geometry} we find that 
\begin{equation}\label{eq:l1l2constraint}
\sinh\frac{l_1}{2}\sinh\frac{l_2}{2} \geq \cosh \frac{l_0}{4}
\end{equation}
and
\begin{equation}\label{eq:thetainl1l2}
\sin \theta = \frac{\cosh\frac{l_0}{4}}{\sinh\frac{l_1}{2}\sinh\frac{l_2}{2}}.
\end{equation}
In view of the explicit form (\ref{dop}) of the operator $\hat D$, we are particularly
interested in the range of the variable conjugate to $\theta$, thus in finding a  
function $\rho$ on Teichm\"uller space satisfying
\begin{equation}
\{\theta,\rho\}_{WP}=1.
\end{equation}
We will for the moment restrict our attention to only half of Teichm\"uller space, corresponding 
to $0 < \theta < \frac{\pi}{2}$, for which we can use $l_1$ and $l_2$ as coordinates 
(with domain given by (\ref{eq:l1l2constraint})). We can find an explicit solution for $\rho(l_1,l_2)$ by 
solving a partial differential equation which we obtain from (\ref{eq:thetainl1l2}) using Wolpert's 
formula (\ref{eq:wolpertformula}),
\begin{eqnarray}
1 =\{\theta,\rho\}_{WP}& = & \cos \theta \left(\frac{\partial \theta}{\partial l_1} 
\frac{\partial \rho}{\partial l_2}-\frac{\partial \theta}{\partial l_2} \frac{\partial \rho}{\partial l_1}\right) \nonumber\\
& = & \frac{\partial (\sin\theta)}{\partial l_1} \frac{\partial \rho}{\partial l_2}-\frac{\partial (\sin\theta)}{\partial l_2} 
\frac{\partial \rho}{\partial l_1} \nonumber\\
& = & \frac{\sin \theta}{2} \left(\coth \frac{l_2}{2} \frac{\partial \rho}{\partial l_1} - \coth 
\frac{l_1}{2} \frac{\partial \rho}{\partial l_2}\right) \label{eq:differentialequationrho1}.
\end{eqnarray}
This equation can be solved using standard techniques for first-order partial differential equations. 
The solution will be unique up to addition of a function of $\theta$, which obviously will
Poisson-commute with $\theta$. To determine $\rho$ uniquely (up to a constant) we require it to 
be antisymmetric in $l_1$ and $l_2$. An uninspiring calculation then leads to
\begin{equation}
\rho = \frac{2}{\sin \theta} \mbox{sc}^{-1} \left(\frac{1}{2}\left(\frac{\cosh\frac{l_1}{2}}{\cosh\frac{l_2}{2}}-\frac{\cosh\frac{l_2}{2}}{\cosh\frac{l_1}{2}}\right)\middle|1-\frac{\sin^2 \theta}{\cosh^2 \frac{l_0}{4}}\right),
\end{equation}
where $\mbox{sc}^{-1}$ is the inverse Jacobi elliptic function \cite{abramowitz_handbook_1965}. 
\begin{figure}
\begin{center}
\centering
\begin{tabular}{cc}
(a) \includegraphics[width=5cm]{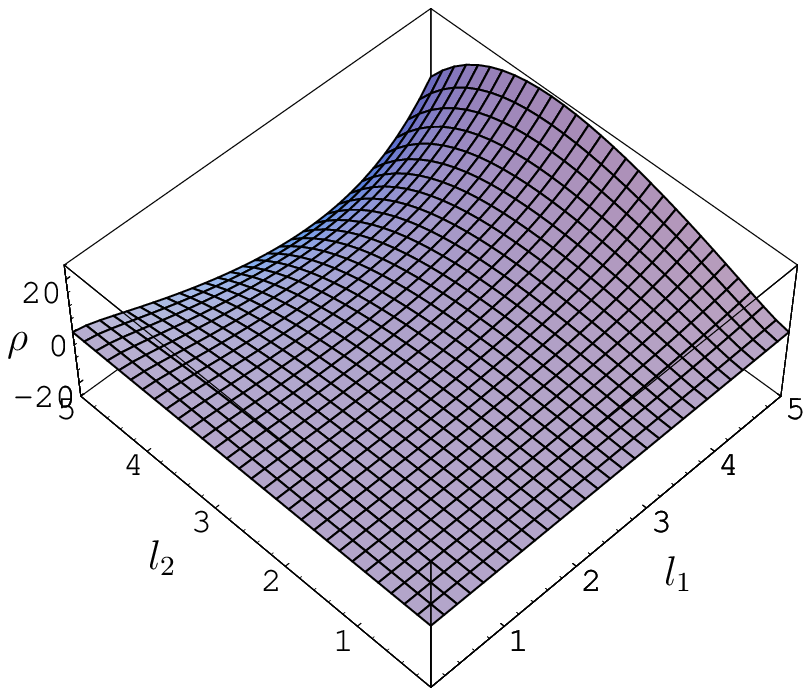} &
(b) \includegraphics[width=4cm]{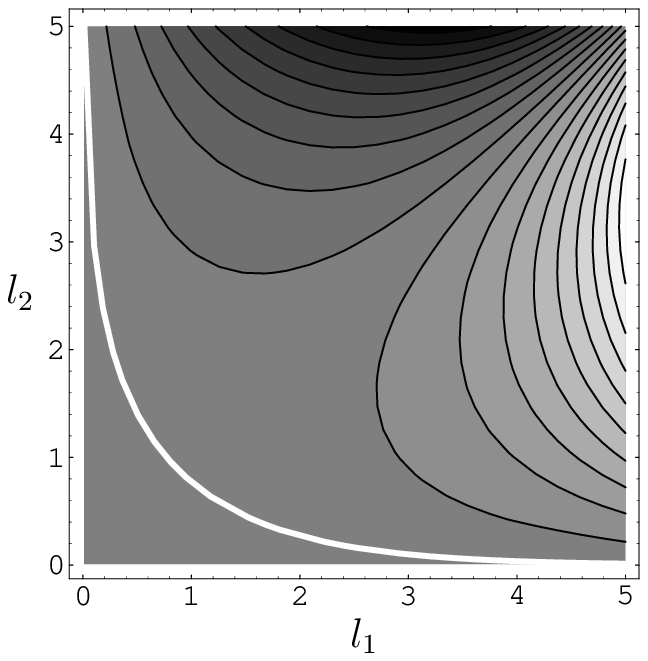} 
\end{tabular}
\caption{(a) Three-dimensional plot and (b) contour plot of $\rho$ as function of $l_1$ and $l_2$ (with $l_0 = 1$). 
The white line in (b) corresponds to $\sinh\frac{l_1}{2}\sinh\frac{l_2}{2}=\cosh\frac{l_0}{4}$. 
Note the antisymmetry with respect to exchange of $l_1$ and $l_2$.
\label{figure:rhoplot}}
\end{center}
\end{figure}
For illustration, we show some Mathematica plots of $\rho$ as function of $l_1$ and $l_2$ for small $l_i$
in Fig.\ \ref{figure:rhoplot}. 
It is not difficult to verify that 
\begin{equation}
\fl\quad\quad \left\{(l_1,l_2)\in \R_{>0}^2 \middle| \sinh \frac{l_1}{2} \sinh \frac{l_2}{2} > \cosh \frac{l_0}{4} \right\} 
\to \left]0,\frac{\pi}{2}\right[ \,\times\, \R : (l_1,l_2) \to (\theta,\rho)
\end{equation}
is smooth and injective. By allowing $\theta$ to take values in $]0,\pi[$, we obtain global coordinates on 
Teichm\"uller space. 

\begin{figure}
\begin{center}
\centering
\begin{tabular}{cc}
(a)\includegraphics[width=4cm]{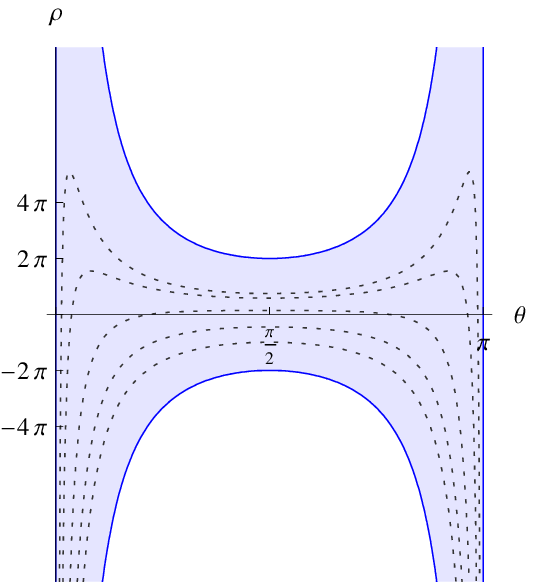} &
(b)\includegraphics[width=5cm]{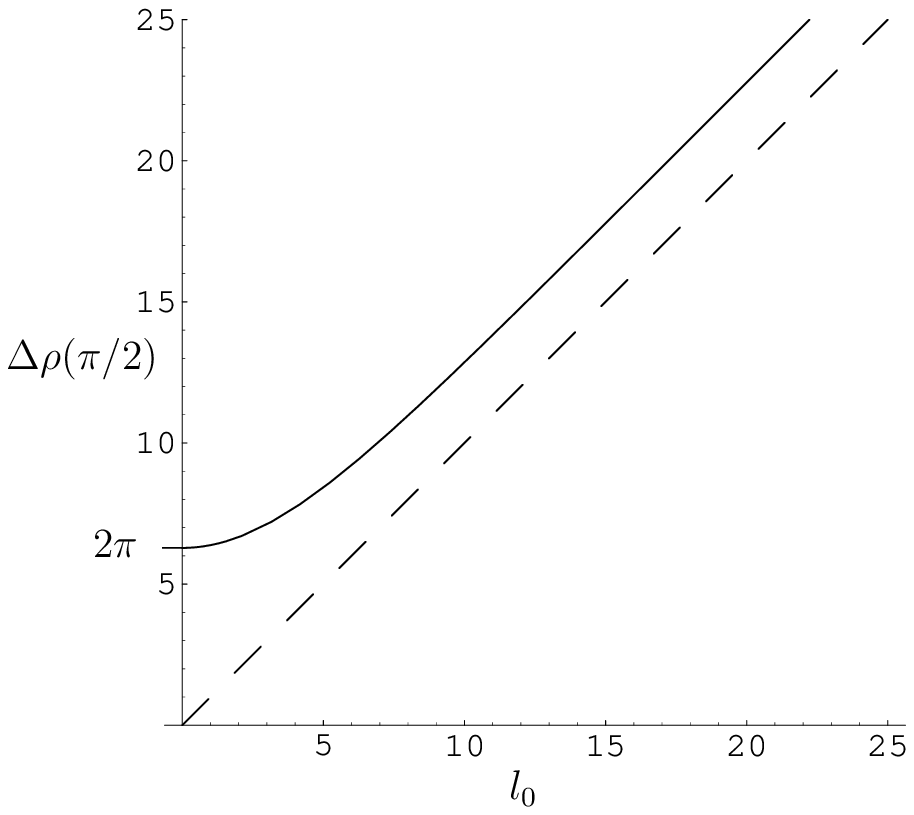} 
\end{tabular}
\caption{(a) The domain of $\rho$ and $\theta$ (shaded area). The dotted curves correspond to 
constant $l_1$. (b) The value of the minimum of $\Delta\rho$, as a function of $l_0$. 
\label{figure:rhophasespace}}
\end{center}
\end{figure}

To find the domain of $\rho$ we note that $x\to\mbox{sc}^{-1}(x|m)$ is a bounded, strictly increasing 
function for fixed $m \in ]-1,1[$. The asymptotic values are $\pm K(m)$ at $x\to\pm \infty$, where $K(m)$ is 
the complete elliptic integral of the first kind \cite{abramowitz_handbook_1965}. Hence, for fixed $\theta$ we 
have $-\frac{1}{2}\Delta\rho_{l_0}(\theta) < \rho < \frac{1}{2}\Delta\rho_{l_0}(\theta)$, where we have defined
\begin{equation}
\Delta\rho_{l_0}(\theta) = \frac{4}{\sin\theta}\ K\left(1-\frac{\sin^2\theta}{\cosh^2\frac{l_0}{4}}\right).
\end{equation}
The function $\Delta\rho_{l_0}(\theta)$ has a minimum at $\theta=\pi/2$, where it assumes the value
\begin{equation}
\Delta\rho_{l_0}\left(\frac{\pi}{2}\right)=4\ K\left(\tanh^2\frac{l_0}{4}\right).
\end{equation}
Computing the minimum as a function of $l_0$ (Fig.\ \ref{figure:rhophasespace}), one observes that it 
starts out at the value $2\pi$ at $l_0=0$ and for increasing $l_0$ converges rapidly 
to $l_0+c$ for a constant $c\approx 2.77$.

We conclude that the separation of the eigenvalues of $\hat{D}_{\alpha_1\alpha_2}$ depends on both
$\theta$ and $l_0$ and is given by 
\begin{equation}
D_{\alpha_1\alpha_2} \in \frac{2\pi} {\Delta\rho_{l_0}(\theta)}\hbar \,\Z ,
\end{equation}
up to a constant which may depend on $\theta$ and $l_0$.
For $\theta$ near $\pi/2$ and $l_0$ small the separation is approximately equal to the Planck length $\hbar$. 
However, the discretization disappears when $\theta\to 0,\pi$ or $l_0\to\infty$.

\begin{figure}
\begin{center}
\centering
\includegraphics[width=6cm]{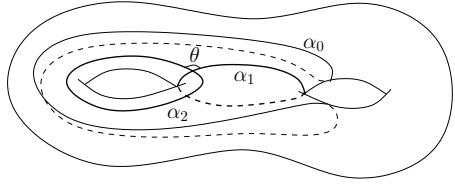}
\caption{A Riemann surface with two simple closed geodesics $\alpha_1$ and $\alpha_2$
intersecting with angle $\theta$. \label{figure:anglegeneralcase}}
\end{center}
\end{figure}

In order to complete our derivation, we still need to show that ``isolating a handle", as we did above 
(c.f. Fig.\ \ref{figure:angleoneholedtorus}), does not constitute any loss of generality.
Let $\Sigma$ be a Riemann surface of any genus $g\geq 2$, and $\alpha_1$ and $\alpha_2$ two 
simple closed geodesics on $\Sigma$ with precisely one intersection (an example with $g=2$
is depicted 
in Fig.\ \ref{figure:anglegeneralcase}). The unique closed geodesic $\alpha_0$ in the homotopy class $[\alpha_1][\alpha_2][\alpha_1]^{-1}[\alpha_2]^{-1}$ is necessarily disjoint from $\alpha_1$ and $\alpha_2$. 
A pair-of-pants decomposition containing $\alpha_0$ and $\alpha_1$ as cuts will then contain one pair of 
pants which has the form of a one-holed torus, as in our previous calculation, the  
only difference being that $l_0$ is no longer an external parameter, but a function on Teichm\"uller space. 
The symplectic structure is given by 
\begin{equation}
\omega_{WP} = \rmd l_0\wedge \rmd \tau_0 + \rmd l_1 \wedge \rmd \tau_1 + \sum_{i=2}^{3g-2}\rmd l_i\wedge \rmd \tau_i,
\end{equation}
with $l_1=l(\alpha_1)$, which can be rewritten as
\begin{equation}\label{eq:thetarhohighergenus}
\omega_{WP} = \rmd l_0\wedge \rmd \tilde{\tau}_0 + \rmd \theta \wedge \rmd \rho + \sum_{i=2}^{3g-2}\rmd l_i\wedge \rmd \tau_i,
\end{equation}
where $\tilde{\tau}_0 = \tau_0 + \Delta\tau_0(l_0,l_1,\tau_1)$ and $\Delta\tau_0$ is a function satisfying 
\begin{equation}
\frac{\partial\Delta\tau_0}{\partial\tau_1} = \frac{\partial \theta}{\partial \tau_1}\frac{\partial \rho}{\partial l_0}-
\frac{\partial \theta}{\partial l_0}\frac{\partial \rho}{\partial \tau_1},\quad
\frac{\partial\Delta\tau_0}{\partial l_1} = \frac{\partial \theta}{\partial l_1}\frac{\partial \rho}{\partial l_0}-
\frac{\partial \theta}{\partial l_0}\frac{\partial \rho}{\partial l_1}.
\end{equation}
One can check that these differential equations are consistent, that is, 
$\partial^2 \Delta\tau_0/\partial l_1\partial\tau_1 = \partial^2 \Delta\tau_0/\partial \tau_1\partial l_1$, 
and therefore can always be solved. We conclude that $\theta,\rho,l_0,\tilde{\tau}_0,l_i,\tau_i$ form a 
new coordinate system for Teichm\"uller space in which $\omega_{WP}$ is given by 
(\ref{eq:thetarhohighergenus}). This means that the spectrum of $\hat{D}_{\alpha_1\alpha_2}$ we found for the 
one-holed torus is valid in this case too. 

A similar argument can be made in the case that the geodesics $\alpha_1$ and $\alpha_2$ do not intersect
on $\Sigma$. 
Recall that the corresponding spacelike distance $D_{\alpha_1\alpha_2}$ 
was the variation of the hyperbolic length 
$h_{\alpha_1\alpha_2}$ of the geodesic connecting them. Also in this case one can always find a pair-of-pants 
decomposition having a particular pair of pants with $\alpha_1$, $\alpha_2$ and a third simple closed 
geodesic $\alpha_0$ as boundary components, and which contains the connecting geodesic. 
Since the geometry of a pair of pants is fully determined by the lengths of its boundary components 
$l_1$, $l_2$ and $l_0$, the length $h_{\alpha_1\alpha_2}$ as a function of the Fenchel-Nielsen coordinates also
depends on $l_1$, $l_2$ and $l_0$ only, and we can write
\begin{equation}
\hat{D}_{\alpha_1\alpha_2} = i\hbar \left(
\frac{\partial h}{\partial l_1} \frac{\partial}{\partial \tau_1} + \frac{\partial h}{\partial l_2} \frac{\partial}{\partial \tau_2} + 
\frac{\partial h}{\partial l_0} \frac{\partial}{\partial \tau_0}\right) .
\end{equation}
Just as in the case of $\hat{L}_{\alpha}$, the Hamiltonian flow is a linear flow in the 
twist parameters and 
therefore the spectrum of $\hat{D}_{\alpha_1\alpha_2}$, for {\it spacelike} distances 
${D}_{\alpha_1\alpha_2}$, is again continuous.

\section{Discussion and conclusion}\label{sec:conclusions}

In this paper, we have identified space- and timelike length variables in 2+1 dimensional
gravity with vanishing cosmological constant. They are given in terms of functions on the
reduced phase space of the theory, obtained in a Chern-Simons formulation of the
three-dimensional Poincar\'e group. Being linear in momenta, the quantization of these
Dirac observables is essentially unique. A study of their eigenvalues in the quantum theory
revealed continuous spectra spanning the entire real line for both space- and timelike
distance operators.\footnote{The length eigenvalues can have either sign because
they came from oriented lengths.} As far as we are aware, this
constitutes the first rigorous derivation of quantum spectra of Dirac length observables
in Lorentzian three-dimensional gravity for genus $g\geq 2$. 

Although our results do not confirm previous investigations in 
\cite{hooft_quantization_1996,freidel_spectra_2003}, which found evidence for a discrete
spectrum for timelike distances, we did come across some discrete aspects in our 
spectral analysis. Although none of the 
length observables we considered were canonically conjugate to an angle, the timelike distance 
$D_{\alpha_1\alpha_2}$ was found to be conjugate to a function with a finite domain. 
However, the size of this domain is not bounded as a function on Teichm\"uller space (the
habitat of the wave functions), which
implies that there is no ``spectral gap" for timelike distances.\footnote{This situation is somewhat
reminiscent of the numerically found properties of the spectrum of the volume operator 
on higher-valence states
in canonical loop quantum gravity in 3+1 dimensions \cite{Brunnemann:2007ca}.}

The discrepancy with previous results may have to do with the fact that
neither of them was based on a complete and consistent quantization of the 
theory on the reduced, physical phase space. The underlying formulations 
are sufficiently different from ours to make a direct comparison difficult. Subtleties with
regard to the implementation of the Hamiltonian constraint \cite{Waelbroeck:1996sg,dittrich_arespectra_2007,rovelli_commentarespectra_2007}
may well play a role. They can be seen as part of a larger issue, present in all but the
simplest systems with gauge symmetry, namely, to what extent quantization and the
imposition of constraints commute \cite{Loll:1990rx,carlip_quantum_2001}. 
Not even for the case of gravity on a spatial torus ($g=1$), whose quantization has 
received a lot of attention
in the physics literature \cite{carlip_quantum_2003} has the question of the equivalence 
or otherwise of different quantizations been settled  completely. Part of the problem is the scarcity of
``observables" which one would like to use to compare physical results.   

The generic presence of quantization ambiguities highlights the fact that the issue of
``fundamental discreteness" can be interpreted in more than one way, depending on
which quantization and operators one applies it to, and therefore may not have a 
unique answer. In the present work, we have focused on the well-defined notion of
investigating the spectra of Dirac observables measuring lengths, obtained in a
``time-less" phase space reduction of three-dimensional
quantum gravity. One could argue that this setting is distinguished, because of the
absence of any choice of time-slicing and the simplicity of the ensuing (Schr\"odinger)
quantization.

The results we have been able to derive come with some
qualifications. Firstly, as already mentioned earlier, the lengths $L_\alpha$ and 
$D_{\alpha_1\alpha_2}$ are not interpretable directly as lengths of curves (or of
distances between such curves) inside the spacetime manifold itself. This happens because
there are no closed geodesics in a nondegenerate solution in the class of geometries
we have been considering (recall that each solution is obtained by making identifications
{\it on a convex open subset} of 3d Minkowski space). 
Nevertheless, they constitute a complete set of length variables ``associated with a solution",
in terms of which any other observable can be expressed. By the same token, we do not
claim that our length variables are directly measurable.\footnote{Of course, physical 
``measurability" is a somewhat academic concept in an unphysical toy model like 
three-dimensional quantum gravity.}

There are related constructions which may yield 
length observables with a more immediate physical interpretation. For example, 
we could consider the length of a path in a particular homotopy class in the limit as it 
approaches the initial singularity, or the lengths of closed geodesics in a
surface of constant cosmological time \cite{benedetti_cosmological_2001}. In either case 
it is difficult to characterize the corresponding functions on Teichm\"uller space explicitly. 
To quantize them one should reformulate the phase space entirely in terms of so-called 
measured laminations. This may be feasible, in the sense that these structures are 
well studied and a lot is known about the relevant symplectic structure 
\cite{benedetti_canonical_2005,bonsante_linear_2005,mcmullen_complex_1998,meusburger_cosmological_2008}. 

Another possibility of constructing physical observables is by enlarging the phase space slightly.  
It is straightforward to include point particles into the model, although there are some subtleties 
which prevent the na\"ive use of a quotient construction to obtain the spacetime. 
The world lines of massive particles define timelike geodesics and one could consider
measuring minimal distances between them. 
An alternative method proposed recently by Meusburger \cite{meusburger_cosmological_2008}
is to define diffeomorphism-invariant observables corresponding to geodesics 
(in this case light-like), but parametrized by the eigentime along the worldline of an observer. 
They are an example of Rovelli's \emph{evolving constants of motion} 
\cite{rovelli_quantum_1990,rovelli_quantum_2004}. 

Note that in our investigation we have only considered length spectra associated with a
subset of curves, namely, particular geodesics (i.e. straight lines) in Minkowski space.
Our construction does not allow for an easy generalization to arbitrary curves.
This, and the peculiar
behaviour we found when analyzing the spectrum of the timelike distance between two
spacelike geodesics in 
the previous section, namely, that the discretization unit of this distance depends on
the relative angle between the geodesics, are an expression of the fact that
the only dynamical degrees of freedom of the theory are of a global nature, and
are captured in a coupled and nonlocal way by various length variables. This is not
a feature one would expect to be present in four dimensions, where the metric does
possess local degrees of freedom. 

The Lorentzian nature of the spacetime was crucial for deriving the results presented here.
If we replaced the Poincar\'e group $ISO(2,1)$ with the Euclidean 
group $ISO(3)$, we would obtain a
theory closely related to the Euclidean lattice gravity model of Ponzano and 
Regge \cite{ooguri_discrete_1991}, whose phase space can be identified with the 
tangent bundle to (a suitable subspace of) the space of flat $SU(2)$-connections 
on $\Sigma$. One can repeat the constructions of Sec.\ \ref{sec:observables} 
in terms of invariant geodesics 
to obtain the analogues of the functions $L_{\alpha}$ and $D_{\alpha_1\alpha_2}$. 
The quantization is completely analogous, with $\hat{L}_{\alpha}$ generating a 
so-called generalized twist flow \cite{goldman_invariant_1986} on the $SU(2)$-equivalent 
of Teichm\"uller space. However, it turns out that this twist flow is periodic with fixed 
period \cite{jeffrey_bohr-sommerfeld_1992}, which implies that the spectrum of 
$\hat{L}_{\alpha}$ will be discretized in units of a fixed minimal length of the order of 
the Planck length. This is in complete 
agreement with results obtained in the loop representation \cite{rovelli_basis_1993}. 

Finally, one may wonder whether any of the techniques we have used can be extended 
to $3+1$ dimensions. An obvious starting point would be a generalization to topological 
field theories with a different gauge group. One such theory, perhaps closest to 
general relativity in $3+1$ dimensions, is BF theory \cite{baez_four-dimensional_1996} 
with gauge group $SO(3,1)$. Since $SO_0(3,1)$ is isomorphic to $PSL(2,\C)$, the 
isometry group of three-dimensional hyperbolic space, one should be able to relate 
some length observables in a flat $3+1$ dimensional spacetime to functions in 
three-dimensional hyperbolic geometry. Further research is needed to determine
whether this is feasible. 
Another connection worthwhile pursuing may be the generalization to $3+1$ dimensions 
of $2+1$ gravity with point particles formulated recently by 't Hooft \cite{hooft_locally_2008}.

\vspace{.8cm}
\noindent {\bfseries {\slshape Acknowledgements.}} RL acknowledges support by 
ENRAGE (European Network on
Random Geometry), a Marie Curie Research Training Network in the
European Community's Sixth Framework Programme, network contract
MRTN-CT-2004-005616, and by the Netherlands Organisation for Scientific Research (NWO) 
under their VICI program.

\appendix
\section{The gauge group}\label{sec:gaugegroup}

We denote the future-preserving Lorentz group in 2+1 dimensions by $SO_0(2,1)$, 
which is precisely the 
identity component of $SO(2,1)$. A basis for its Lie algebra $\mathfrak{so}(2,1)$ is given by
\begin{equation}\label{eq:basisso21}
\fl\quad\quad \begin{array}{ccc}
J_0^{\mathfrak{so}} = \left( \begin{array}{ccc} 0 & 0 & 0 \\ 0 & 0 & 1 \\ 0 & -1 & 0 \end{array} \right) &
J_1^{\mathfrak{so}} = \left( \begin{array}{ccc} 0 & 0 & -1 \\ 0 & 0 & 0 \\ -1 & 0 & 0 \end{array} \right) &
J_2^{\mathfrak{so}} = \left( \begin{array}{ccc} 0 & -1 & 0 \\ -1 & 0 & 0 \\ 0 & 0 & 0 \end{array} \right). 
\end{array}
\end{equation}
satisfying the commutation relations $[J_{a},J_{b}]={\epsilon_{a b}}^{c}J_{c}$, where the totally antisymmetric
$\epsilon$-tensor is defined by $\epsilon_{012}=-\epsilon^{012}=1$, and indices are raised and lowered
with the metric $\eta_{ab}={\rm diag}(-1,1,1)$. The generators $J_a^{\mathfrak{so}}$
form an orthonormal basis with respect to the indefinite bilinear form
\begin{equation}
B^{\mathfrak{so}}(X,Y)=\frac{1}{2}\Tr(X Y),\;\;\;\;\; 
B^{\mathfrak{so}}(J_a^{\mathfrak{so}},J_b^{\mathfrak{so}})=\eta_{ab}.
\end{equation}
We will often use the isomorphism $SO_0(2,1)\cong PSL(2,\R)=SL(2,\R)/\{I,-I\}$. To make the isomorphism 
explicit we choose the specific basis 
\begin{equation}\label{eq:basissl2}
\begin{array}{ccc}
J_0^{\mathfrak{sl}} = \frac{1}{2} \left( \begin{array}{cc} 0 & -1 \\ 1 & 0 \end{array} \right) &
J_1^{\mathfrak{sl}} = \frac{1}{2} \left( \begin{array}{cc} 1 & 0 \\ 0 & -1 \end{array} \right) &
J_2^{\mathfrak{sl}} = \frac{1}{2} \left( \begin{array}{cc} 0 & 1 \\ 1 & 0 \end{array} \right)
\end{array}
\end{equation}
for the Lie algebra $\mathfrak{sl}(2,\R)$. The generators 
satisfy identical commutation relations and are orthonormal with respect to the bilinear form
\begin{equation}\label{eq:bilinearformsl}
B^{\mathfrak{sl}}(X,Y)=2\Tr(XY), \;\;\;\;\; 
B^{\mathfrak{sl}}(J_a^{\mathfrak{sl}},J_b^{\mathfrak{sl}})=\eta_{ab}.
\end{equation}
In fact, $SO_0(2,1)$ emerges as the adjoint representation of $PSL(2,\R)$ on $\mathfrak{sl}(2,\R)$, 
when written in the basis (\ref{eq:basissl2}). 

The gauge group of 2+1 gravity is the 2+1 dimensional Poincar\'e group $ISO_0(2,1)$, which is most easily 
characterized as the semi-direct product of $SO_0(2,1)$ and the abelian group $\R^3$, where the action of 
$SO_0(2,1)$ on $\R^3$ is the fundamental one, namely,
\begin{equation}
(g_1,X_1)\cdot(g_2,X_2) = (g_1 g_2,X_1 + g_1 X_2).
\end{equation}
In terms of the basis (\ref{eq:basisso21}), we can identify $\mathfrak{so}(2,1)$ with $\R^3$, where 
the action of $SO_0(2,1)$ now becomes the adjoint action,
\begin{equation}
(g_1,X_1)\cdot(g_2,X_2) = (g_1 g_2,X_1 + \Ad(g_1) X_2).
\end{equation}
In this way we identify $ISO_0(2,1)$ with $SO_0(2,1) \ltimes \mathfrak{so}(2,1)$, which is again 
isomorphic to $PSL(2,\R)\ltimes \mathfrak{sl}(2,\R)$.

Furthermore, for any Lie group $G$ its semi-direct product group 
$G \ltimes \mathfrak{g}$ is isomorphic to its \emph{tangent group} $TG$, which is defined by taking as 
multiplication the  tangent map to the 
multiplication map $G \times G \to G$. Explicitly, the isomorphism, which elsewhere we often use implicitly, 
identifies the elements of the Lie algebra $\mathfrak{g}$ with the right-invariant vector fields on 
$G$. -- To summarize, we have the following chain of isomorphisms
\begin{eqnarray}
ISO_0(2,1) &\cong& SO_0(2,1)\ltimes\mathfrak{so}(2,1) \cong PSL(2,\R)\ltimes\mathfrak{sl}(2,\R)\nonumber\\
&\cong& T\,SO_0(2,1) \cong T\,PSL(2,\R).
\end{eqnarray}

Finally, a nondegenerate bilinear form on $\mathfrak{g}$ gives rise to a natural bilinear form on the 
Lie algebra of $TG$ by taking its derivative. For the case at hand, we obtain a natural nondegenerate bilinear 
form on $\mathfrak{iso}(2,1)\cong\mathfrak{so}(2,1)\times\mathfrak{so}(2,1)$ by defining
\begin{equation}
B\left((X_1,Y_1),(X_2,Y_2)\right) = B^{\mathfrak{so}}(X_1,Y_2)+B^{\mathfrak{so}}(Y_1,X_2).
\end{equation}

\section{Some group theory}\label{sec:grouptheory}

For a Lie group $G$ and its associated Lie algebra $\mathfrak{g}$, we denote the left and right multiplication maps by $l_g,r_g:G\to G$. The conjugation map $C_g: x\to gxg^{-1}$ is an 
isomorphism of $G$ to itself and its tangent map at the origin $\Ad(g)=T_eC_g$ is the adjoint representation acting on $\mathfrak{g}$. Suppose $B$ is a nondegenerate invariant 
(pseudo-) metric on $G$, i.e. $B:\mathfrak{g}\times\mathfrak{g}\to\R$ is a nondegenerate 
bilinear form invariant under adjoint transformations. We denote by $\tilde{B}$ the associated 
map $\mathfrak{g}\to\mathfrak{g}^*$.

To a differentiable function $f:G\to\R$ we can associate a natural map 
$\hat{\xi}_f:G\to \mathfrak{g}^*$, which is the right translation of the derivative of $f$ to the 
Lie algebra,
\begin{equation}\label{eq:xihatdef}
\hat{\xi}_f(g) = (T_er_g)^* \rmd f(g).
\end{equation}
Using the metric $B$, we can define the \emph{variation} 
$\xi_f = \tilde{B}^{-1} \circ \hat{\xi}_f : G \to \mathfrak{g}$ of $f$ \cite{goldman_invariant_1986}. Equivalently,
\begin{equation}\label{eq:xidef}
B(\xi_f(g),X)= \left.\frac{\rmd }{\rmd t}\right|_{t=0} f(\exp(tX)g)
\end{equation}
for $X\in\mathfrak{g}$. Here $\exp:\mathfrak{g}\to G$ is the standard exponential map for Lie groups.

From now on will we assume $f$ to be a \emph{class function}, i.e. a function which is invariant under 
conjugation, $C_g^*f=f$ for all $g\in G$. Using the well-known identity $C_h \circ \exp = \exp \circ \Ad(h)$, we find that
\begin{equation}
\xi_f(hgh^{-1})=\Ad(h)\xi_f(g).
\end{equation}
Putting $h=g$ we see that $\xi_f(g)$ is invariant under $\Ad(g)$,
\begin{equation}\label{eq:adginvariant}
\Ad(g)\xi_f(g)=\xi_f(g).
\end{equation}

Consider now the specific case $G=PSL(2,\R)$ with the metric as in (\ref{eq:bilinearformsl}). 
For hyperbolic elements $g\in G_{hyp}=\{g|\Tr(g)>2\}\subset G$, we define the 
\emph{hyperbolic length} $l(g)$ of $g$ by
\begin{equation}\label{eq:definitionlg}
\Tr(g)=2 \cosh( l(g)/2 ).
\end{equation}
Due to the cyclicity of the trace, $l$ is a class function, whose variation we can compute in
a straightforward manner. Applying relation (\ref{eq:xidef}), 
we find for diagonal elements $g$ that $\xi_l(g)=\mathrm{diag}(1/2,-1/2)$. 
For general elements $g\in G_{hyp}$ which are diagonalized by $h\in G$, we 
have $\xi_l(g)=\Ad(h)\mathrm{diag}(1/2,-1/2)$. In particular, $\xi_l(g)$ is spacelike,
of unit norm and the group element can be written as \cite{goldman_invariant_1986}
\begin{equation}\label{eq:explgxilg}
g=\exp(l(g)\xi_l(g)).
\end{equation}
If desired, these two conditions can be taken as the definition of $l(g)$ and $\xi_l(g)$, because the exponential map is a bijection from $\{X\in\mathfrak{sl}(2,\R)|B(X,X)>0\}$ to $G_{hyp}$. 
One can write the exponential in (\ref{eq:explgxilg}) explicitly as
\begin{equation}\label{eq:explicitexp}
g = \cosh\frac{l(g)}{2} \ 1 + \left( 2 \sinh\frac{l(g)}{2}\right) \xi_l(g)
\end{equation}
and
\begin{equation}\label{eq:explicitexpad}
\fl\quad\quad\quad\Ad(g) = \cosh l(g)\ 1 + (1-\cosh l(g))\xi_l(g)B(\xi_l(g),\cdot) + \sinh l(g) [\xi_l(g),\cdot].
\end{equation}

For elliptic elements we have a similar construction. We define $\theta(g)\in[0,2\pi[$ by 
$\Tr(g)=2\cos(\theta/2)$. In that case, $\xi_{\theta}(g)$ is timelike, of unit norm and
the group element can be written as
\begin{equation}
g=\exp(\theta(g)\xi_{\theta}(g)).
\end{equation}

\section{Hyperbolic geometry}\label{sec:hyperbolic geometry}

In this appendix we will briefly review some notions of hyperbolic geometry, which are used 
in the main text (see, for example, \cite{imayoshi_introduction_1992} for details and proofs). 
For our purposes, ``hyperbolic geometry" will mean the study of Riemann surfaces and geometric constructions on them. A Riemann surface is a two-dimensional real manifold with a complex 
structure. In this article we consider only compact oriented Riemann surfaces, 
which are topologically classified by their genus $g$. We will be concerned only with 
the case $g\geq 2$. 

Let $\Sigma$ be a surface of genus $g\geq 2$. Its fundamental group is generated by a set of $2g$ homotopy classes of closed curves $\{a_i,b_i\}_{i=1,\ldots,g}$ satisfying the relation
\begin{equation}
\prod_{i=1}^g a_i b_i a_i^{-1} b_i^{-1} = 1.
\end{equation}
The space of inequivalent complex structures on $\Sigma$ is called the \emph{moduli space} 
$\mathcal{M}_g$ and depends only on the genus $g$. In the following we will consider a 
slightly larger space, the \emph{Teichm\"uller space} $\mathcal{T}_g$, which is the 
universal covering of $\mathcal{M}_g$. One can define Teichm\"uller space as the space 
of equivalence classes of \emph{marked} Riemann surfaces. ``Marked'' implies that 
we pick a complex structure {\it and} a distinguished set of generators for the 
fundamental group. 
Equivalently, we identify two complex structures on $\Sigma$ if they are related by a 
biholomorphism homotopic to the identity map. The moduli space can be obtained 
from Teichm\"uller space by taking the quotient with respect to the \emph{mapping class group},
\begin{equation}
\mathcal{M}_g = \mathcal{T}_g/\mathcal{MCG}.
\end{equation} 
It is well known that elements of the space of complex structures on $\Sigma$ are in one-to-one 
correspondence with conformally inequivalent metrics on $\Sigma$. Moreover, for the 
case $g\geq 2$, every conformal equivalence class contains a unique 
\emph{hyperbolic metric}, i.e. a metric with constant curvature $-1$. Consequently, we 
can identify Teichm\"uller space with the space of hyperbolic metrics on $\Sigma$ 
modulo diffeomorphisms connected to the identity.

Next, consider a surface $\Sigma$ with a specific complex structure. As a consequence 
of the well-known uniformization theorem, the universal cover of $\Sigma$ is the complex 
upper half-plane 
$\mathbb{H}$. Any element of the fundamental group of $\Sigma$ therefore corresponds 
to an automorphism of $\mathbb{H}$. The automorphism group of $\mathbb{H}$ is easily 
seen to be equal to $PSL(2,\R)$, acting according to
\begin{equation}\label{eq:actionsl2ronh}
\left(\begin{array}{cc} a & b \\ c & d \end{array}\right): z \to \frac{a z +b}{c z+d}.
\end{equation}
An element $g\in PSL(2,\R)$ which is not equal to the identity is said to be 
\emph{hyperbolic} if $|\Tr(g)|>2$, \emph{elliptic} if $|\Tr(g)|<2$ and \emph{parabolic} 
if $\Tr(g)=2$. Under the iden\-ti\-fication $PSL(2,\R)\cong SO_0(2,1)$ these correspond to 
boosts, rotations and lightlike transformations respectively. On $\mathbb{H}$, they can be
characterized as those transformations which leave fixed two, no and one 
points on the boundary respectively.

The set $\Gamma$ of automorphisms corresponding to the elements of the fundamental 
group is called a \emph{Fuchsian model} of $\Sigma$. We can write $\Sigma$ as the quotient
\begin{equation}
\Sigma = \mathbb{H}/\Gamma.
\end{equation}
Since $\Sigma$ is a smooth manifold, $\Gamma$ must act properly discontinuously on 
$\mathbb{H}$, which is equivalent to $\Gamma$ being a \emph{Fuchsian group},
that is, a discrete subgroup of $PSL(2,\R)$. 
Such a group necessarily consists only of hyperbolic elements (and the identity). 
It turns out that any representation of the fundamental group $\pi_1$ as a Fuchsian 
group in $PSL(2,\R)$ arises as the Fuchsian model of a complex structure. 
We therefore can identify Teichm\"uller space as
\begin{equation}
\mathcal{T}_g = \mathrm{Hom}_0(\pi_1(\Sigma),PSL(2,\R))/PSL(2,\R),
\end{equation} 
where the subscript 0 means that we restrict to injective homomorphisms which have a 
Fuchsian group as image.

\begin{figure}[htbp]
\begin{center}
\begin{tabular}{cc}
(a) \includegraphics[width=3cm]{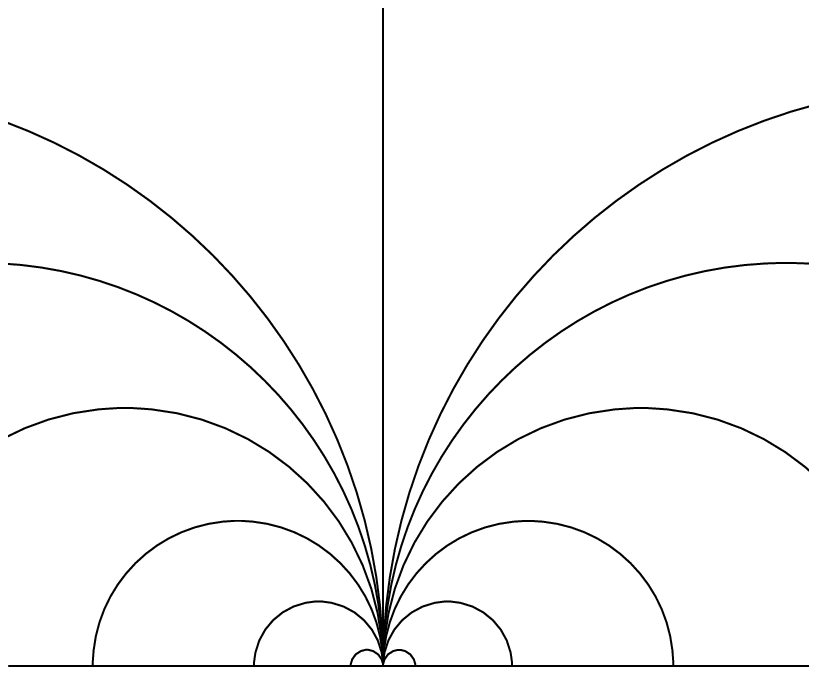}&
(b) \includegraphics[width=3cm]{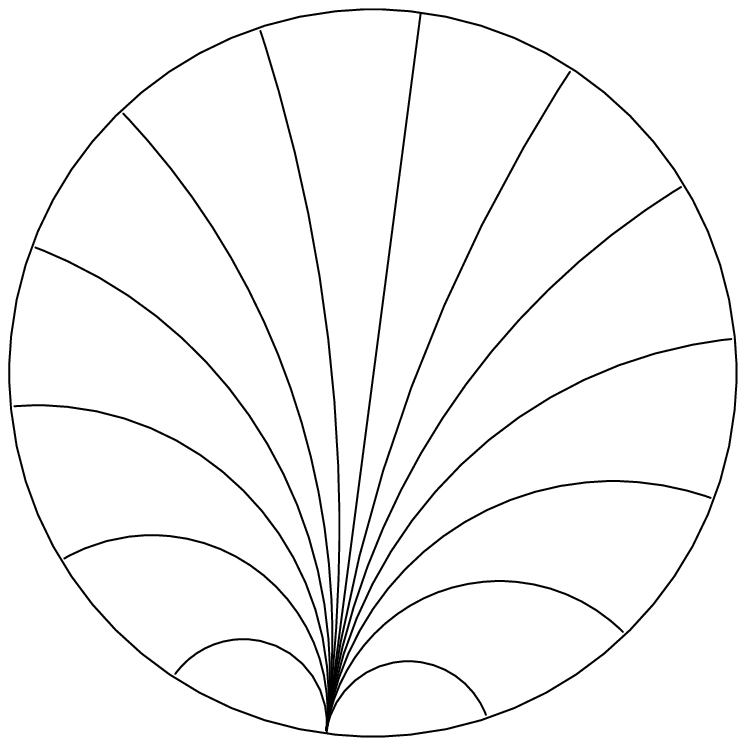}
\end{tabular}
\caption{Some geodesics in (a) the complex upper-half plane $\mathbb{H}$ and (b) 
the Poincar\'e disc $\Delta$.\label{figure:geodesics}}
\end{center}
\end{figure}

To establish the geometric properties of a Riemann surface it suffices to know the 
geometry of its universal covering $\mathbb{H}$. The hyperbolic metric on $\mathbb{H}$ 
corresponding to its complex structure is the Poincar\'e metric 
$\rmd s^2_{\mathbb{H}} = |\rmd z|^2/\mathrm{Im}^2z$. The geodesics with respect to this metric 
are half circles centered on the real axis (see Fig.\ \ref{figure:geodesics}). Sometimes 
it is convenient to use an equivalent model of the hyperbolic plane, the Poincar\'e 
disc $\Delta$ with the Poincar\'e metric
\begin{equation}
\rmd s^2_{\Delta} = \frac{4|\rmd z|^2}{(1-|z|^2)^2}.
\end{equation}
Now the geodesics are circle arcs perpendicular to the boundary (see Fig.\ \ref{figure:geodesics}).

\begin{figure}[htbp]
\begin{center}
\includegraphics[width=5cm]{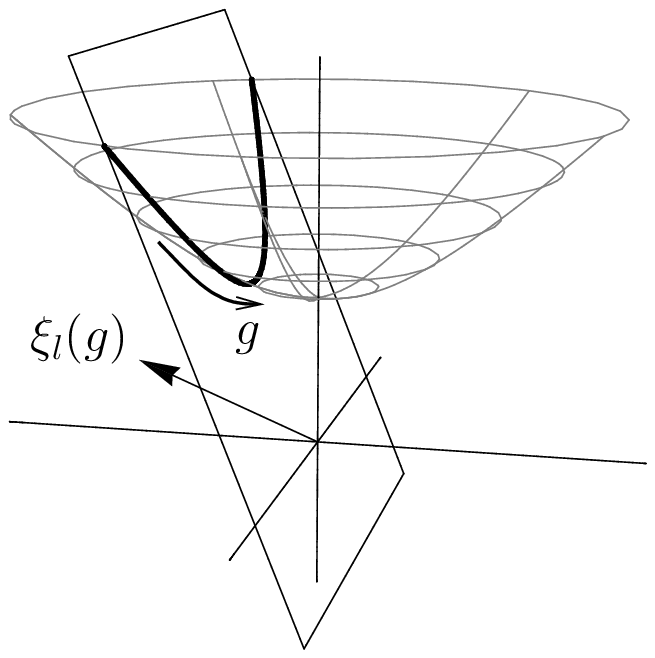}
\caption{The hyperboloid model. \label{figure:hyperboloid}}
\end{center}
\end{figure}

Yet another representation of the hyperbolic plane, which makes the relation to 2+1 
dimensional spacetime most transparent, is the \emph{hyperboloid model}. 
It is defined as the unit hyperboloid
\begin{equation}
H_1 = \{X \in \R^3 | X\cdot X=-1,X^0>0 \}
\end{equation}
in three-dimensional Minkowski space, as depicted in Fig.\ \ref{figure:hyperboloid}, with the 
induced metric. The geodesics are given by the intersections of $H_1$ 
with two-dimensional planes through the origin. An element $g\in PSL(2,\R)\cong SO_0(2,1)$ 
acts on $H_1$ by Lorentz transformation (the adjoint representation $\Ad(g)$ of $g$, 
if we identify Minkowski space with $\mathfrak{sl}(2,\R)$). If $g$ is hyperbolic, it 
corresponds to a boost in Minkowski space. Such a boost leaves a unique plane through
the origin invariant 
and thus $g$ determines a unique invariant geodesic in $H_1$. 
Note that $\xi_l(g)$ is the normal to this plane, since it is invariant under $\Ad(g)$.   

\begin{figure}[htbp]
\begin{center}
\begin{minipage}[c]{0.3\linewidth}
\centering
\includegraphics[width=3cm]{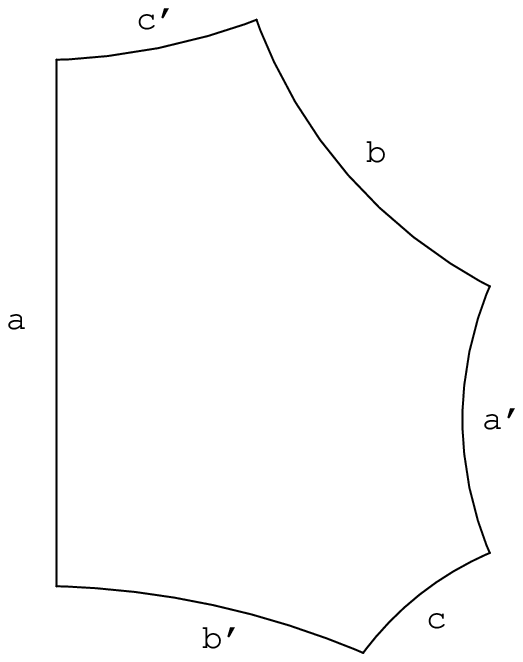}\\
(a)
\end{minipage}
\begin{minipage}[c]{0.3\linewidth}
\centering
\includegraphics[width=3cm]{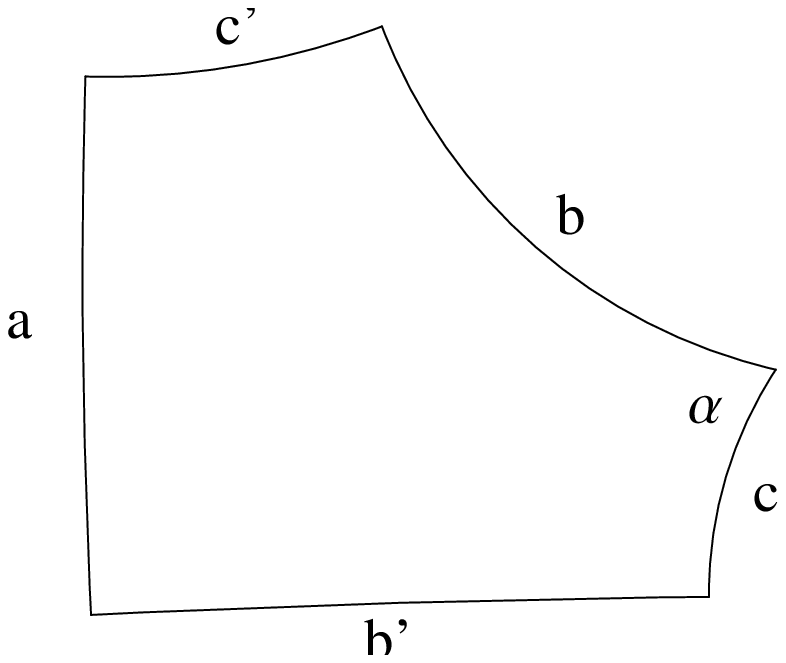}\\
(b)
\end{minipage}
\begin{minipage}[c]{0.3\linewidth}
\centering
\includegraphics[width=3cm]{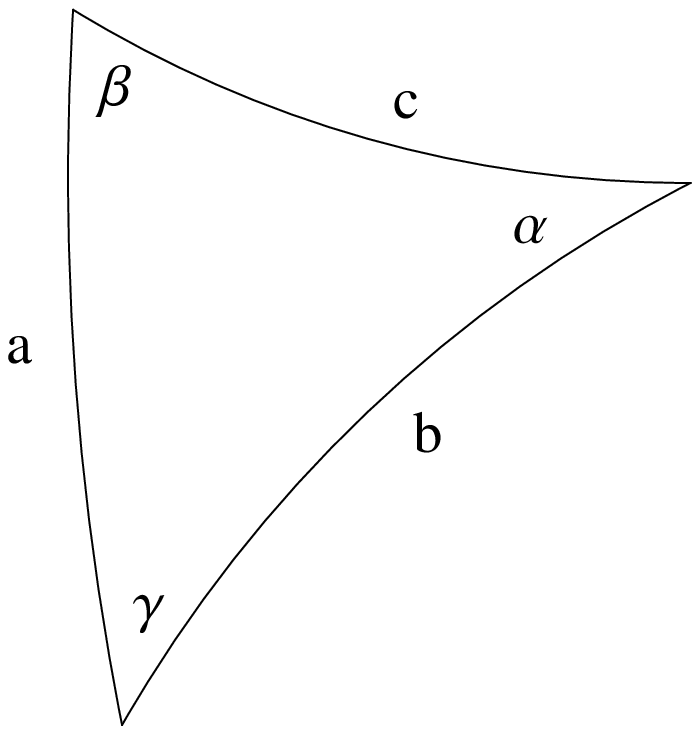}\\
(c)
\end{minipage}
\caption{Examples of hyperbolic polygons. All unlabelled angles are right angles.\label{figure:polygons}}
\end{center}
\end{figure}

Trigonometry can be developed in the hyperbolic plane in analogy with the Euclidean case. 
We state here some trigonometric relations \cite{fenchel_elementary_1989} for 
hyperbolic polygons, which are used in the main text. Referring to the notation of 
Fig.\ \ref{figure:polygons}, they are
\begin{itemize}
\item[(a)] Given any three numbers $a,b,c \in \R_{>0}$ there exists a unique
convex right-angled hexagon with alternating sides of length $a$, $b$ and $c$.
The lengths of the sides satisfy
the relations
\begin{equation}
\frac{\sinh a}{\sinh a'} = \frac{\sinh b}{\sinh b'}=\frac{\sinh c}{\sinh c'},
\end{equation}
\begin{equation}
\cosh a' = \frac{\cosh b \cosh c + \cosh a}{\sinh b \sinh c}.
\end{equation}
Analogous relations hold for $b'$ and $c'$.
\item[(b)] A pentagon with four right angles and a remaining angle $\alpha$ satisfies
\begin{equation}\label{eq:pentagonequality1}
\frac{\sinh a}{\sin \alpha}=\frac{\cosh b}{\sinh b'}=\frac{\cosh c}{\sinh c'},
\end{equation}
\begin{equation}\label{eq:pentagonequality2}
\cosh a = \sinh b \sinh c - \cosh b \cosh c \cos \alpha.
\end{equation}
\item[(c)] A triangle with arbitrary angles satisfies
\begin{equation}
\frac{\sinh a}{\sin \alpha}=\frac{\sinh b}{\sin \beta}=\frac{\sinh c}{\sin
\gamma},
\end{equation}
\begin{equation}
\cosh a = \cosh b \cosh c - \sinh b \sinh c \cos \alpha.
\end{equation}
\end{itemize}

We saw above that a hyperbolic element $g\in PSL(2,\R)$ leaves exactly two points on the 
boundary of the complex upper-half plane $\mathbb{H}$ fixed, which implies that
there is a unique geodesic in $\mathbb{H}$ invariant under $g$. 
The geodesic is translated along itself
by a hyperbolic distance $l(g)$, which is related to the trace of 
$g$ through eq.\ (\ref{eq:definitionlg}). If $g$ is an element of the Fuchsian model of $\Sigma$ corresponding to a homotopy class $\alpha$, this geodesic projects to the unique closed 
geodesic in $\alpha$ with length given by $l(g)$.

Closed geodesics on the Riemann surface are associated with a convenient set of 
coordinates on Teichm\"uller space, known as the \emph{Fenchel-Nielsen coordinates}. 
Given a Riemann surface $\Sigma$ of genus $g\geq2$, one can always find a set of 
$3g-3$ mutually disconnected 
simple (that is, non-selfintersecting) closed geodesics 
$\{\gamma_i\}$. Cutting the surface along these geodesics results in a
decomposition of  $\Sigma$ into $2g-2$ pairs of pants, each one a genus-0 Riemann surface 
with three geodesic boundary components. 
It is not hard to show (using the trigonometric identity (a) above) that the complex structure 
on a pair of pants is completely determined by the lengths of its boundary components. 
In order to fix the complex structure on $\Sigma$ we therefore need to fix the lengths $l_i$ of the 
geodesics $\gamma_i$ and the way we re-glue the pairs of pants. To quantify the latter 
we introduce the so-called twist parameters $\tau_i$, which measure the distance between 
particular distinguished points on $\gamma_i$. It turns out that both types of variables
taken together,
\begin{equation}
(l_i,\tau_i)\in (\R_{>0} \times \R)^{3g-3},
\end{equation}
form a global set of coordinates on Teichm\"uller space. In particular, this implies that
\begin{equation}
\dim \mathcal{T}_g = 6g-6.
\end{equation}

The Weil-Petersson symplectic structure (see \ref{sec:weilpetersson}) takes on a 
particularly simple form in terms of the Fenchel-Nielsen 
coordinates \cite{imayoshi_introduction_1992}, namely,
\begin{equation}\label{eq:weilpeterssoninfenchelnielsen}
\omega_{WP} = \sum_{i=1}^{3g-3} \rmd l_i \wedge \rmd \tau_i.
\end{equation}
For two closed geodesics $\alpha$ and $\beta$, the Poisson bracket of their 
associated lengths variables is given by Wolpert's formula,
\begin{equation}\label{eq:wolpertformula}
\{l_{\alpha},l_{\beta}\}_{WP} = \sum_{p\in \alpha\sharp\beta} \cos \theta_p,
\end{equation}
where the sum runs over the intersection points $p$ and $\theta_p$ is the angle between 
$\alpha$ and $\beta$ at $p$. Note that we could have derived this formula by combining 
(\ref{eq:xiproductangle}) with (\ref{eq:generalizedweilpeterssonbracket}) (see also \cite{goldman_invariant_1986}).

\section{Generalized Weil-Petersson symplectic structure}\label{sec:weilpetersson}

For a compact oriented surface $\Sigma$ of genus $g>1$,  we are interested in 
homomorphisms from its fundamental group $\pi_1$ to a Lie group
$G$. More specifically, we want to consider the space $\mbox{Hom}(\pi_1,G)/G$, 
where $G$ acts on $\mbox{Hom}(\pi_1,G)$ by overall conjugation. For a
homomorphism $\phi: \pi_1
\rightarrow G$, we will denote its equivalence class by $[\phi] \in \mbox{Hom}(\pi_1,G)/G$.

If $G$ possesses a (pseudo-)metric, i.e. a nondegenerate bilinear form $B$ on its Lie algebra
 $\mathfrak{g}$, (a suitable open subset of) the space $\mbox{Hom}(\pi_1,G)/G$ can be given 
a canonical symplectic structure $\omega_G$, known as the \emph{generalized 
Weil-Petersson symplectic structure}. Without giving any details of the construction, 
which involves homology, we will simply state the main result of \cite{goldman_symplectic_1984}. 
For a class function $f$ on $G$ (see \ref{sec:grouptheory}) and a closed curve 
$\alpha$ in $\Sigma$, define
\begin{equation}
f_{\alpha} : \mbox{Hom}(\pi_1,G)/G \rightarrow \R : [\phi] \rightarrow
f(\phi([\alpha])).
\end{equation}
Given two such functions, $f$ and $f'$, and two closed curves $\alpha$ and $\alpha'$, 
the Poisson bracket of $f_{\alpha}$ and $f'_{\alpha'}$ turns out to be
\begin{equation}\label{eq:generalizedweilpeterssonbracket}
\{f_{\alpha},f'_{\alpha'}\}_G([\phi]) = \sum_{p \in \alpha \sharp \alpha'} 
\varepsilon(p;\alpha,\alpha')
B(\xi_f(\phi(\alpha_p)),\xi_{f'}(\phi(\alpha'_p))),
\end{equation}
where $\alpha \sharp \alpha'$ is the set of intersections in $\Sigma$. The discrete variable
$\varepsilon(p;\alpha,\alpha')=\pm 1$ depends on the orientation of the
intersection, $\alpha_p$ is just the curve $\alpha$ but with base point specified to be $p$, 
and $\xi_f : G \to \mathfrak{g}$ is the variation of $f$ as defined in \ref{sec:grouptheory}.

The above can also be applied to the tangent group $TG$, which was introduced in 
\ref{sec:gaugegroup}, together with the metric
\begin{equation}
B_{TG}((X_1,Y_1),(X_2,Y_2)) = B(X_1,Y_2)+B(Y_1,X_2),
\end{equation}
which is essentially the derivative of $B$. We will now show that the generalized 
Weil-Petersson symplectic form $\omega_{TG}$ for the group $TG$ is the tangent 
symplectic form corresponding to $\omega_{G}$. Instead of $TG$ we will use the 
semi-direct product $G\ltimes\mathfrak{g}$, which is isomorphic to $TG$ by right 
translation (see \ref{sec:gaugegroup}).

Let $f:G\to\R$ be a class function on $G$ (see \ref{sec:grouptheory}). We associate to $f$ 
two class functions on $G\ltimes\mathfrak{g}$, its trivial extension $\hat{f}=f\circ\pi_G$ and its
variation $F:(g,X)\to B(\xi_f(g),X)$. To compute the Poisson brackets 
(\ref{eq:generalizedweilpeterssonbracket}) we will need the variations of both $\hat{f}$ and 
$F$. The variation of $\hat{f}$ is easily seen to be given by 
$\xi_{\hat{f}}(g,X) = (0,\xi_f(g))\in\mathfrak{g}\times\mathfrak{g}$. The variation of $F$ is a bit trickier. 
From definition (\ref{eq:xihatdef}) we have 
\begin{eqnarray}
\fl\quad\quad B_{TG}(\xi_F(g,X),(Y,Z)) &=& \rmd F( (g,X), T_e r_{(g,X)} (Y,Z))\nonumber\\
&=& \left.\frac{\rmd }{\rmd s}\right|_{s=0} F\left( \exp(sY)g,X + s Z + s[Y,X])\right)\nonumber\\
&=& \left.\frac{\rmd }{\rmd t}\right|_{t=0} \left.\frac{\rmd }{\rmd s}\right|_{s=0} f\left(\exp(tX+tsZ+ts[Y,X])\exp(sY)g\right)\nonumber\\
&=& \left.\frac{\rmd }{\rmd t}\right|_{t=0} \left.\frac{\rmd }{\rmd s}\right|_{s=0} f\left(\exp(sY)\exp(tX+tsZ)g\right)\nonumber\\
&=& B(\xi_f(g),Z) + \left.\frac{\rmd }{\rmd t}\right|_{t=0}B(\xi_f(\exp(tX)g),Y).
\end{eqnarray}
Hence
\begin{equation}
\xi_F(g,X)=\left(\xi_f(g),\left.\frac{\rmd }{\rmd t}\right|_{t=0}\xi_f(\exp(tX)g)\right).
\end{equation}
Applying the variations to two class functions $f$ and $f'$ we get
\begin{eqnarray}
B_{TG}(\xi_{\hat{f}}(g,X),\xi_{\hat{f}'}(h,Y)) = 0\nonumber\\
B_{TG}(\xi_{F}(g,X),\xi_{\hat{f}'}(h,Y)) = B(\xi_f(g),\xi_{f'}(h))\\
B_{TG}(\xi_{F}(g,X),\xi_{F'}(h,Y)) = \left.\frac{\rmd }{\rmd t}\right|_{t=0} B(\xi_f(\exp(tX)g),\xi_{f'}(\exp(tY)h))\nonumber.
\end{eqnarray}
It now follows from (\ref{eq:generalizedweilpeterssonbracket}) that 
\begin{eqnarray}
\{\hat{f}_{\alpha},\hat{f}'_{\alpha'}\}_{TG}=0,\nonumber\\
\{F_{\alpha},\hat{f}'_{\alpha'}\}_{TG} = \{f_{\alpha},f'_{\alpha'}\}_G \circ \pi_G, \\
\{F_{\alpha},F'_{\alpha'}\}_{TG} = \rmd \{f_{\alpha},f'_{\alpha'}\}_G,\nonumber
\end{eqnarray}
which is precisely the structure (\ref{eq:tangentpoissonbracket}) we expect for the Poisson brackets corresponding to the tangent symplectic structure.

\section*{References}
\bibliographystyle{iopart-numX}
\bibliography{ref}

\end{document}